\documentclass[a4paper,11pt]{article}

\usepackage{a4wide}
\usepackage[english]{babel}

\usepackage{amsfonts}
\usepackage{amsmath}
\usepackage{graphicx}
\usepackage{caption}
\usepackage[colorlinks,bookmarks=true]{hyperref}
\usepackage{amsthm}
\usepackage{enumitem}

\newtheorem{lemma}{Lemma}[section]

\newtheorem{theorem}{Theorem}[section]
\newtheorem{prop}{Proposition}[section]

\newtheorem{assumption}{Assumption}[section]
\newtheorem{rem}{Remark}[section]

\newtheorem{corollary}{Corollary}[section]

\title{The characteristic function of rough Heston models}

\author{Omar El Euch\\ CMAP, \'Ecole Polytechnique Paris \\ omar.el-euch@polytechnique.edu\\$~~$\\
Mathieu Rosenbaum\\ CMAP, \'Ecole Polytechnique Paris\\
  mathieu.rosenbaum@polytechnique.edu}

\begin{document}

\maketitle

\begin{abstract}
\noindent It has been recently shown that rough volatility models, where the volatility is driven by a fractional Brownian motion with small Hurst parameter,
provide very relevant dynamics in order to reproduce the behavior of both historical and implied volatilities. However, due to the non-Markovian nature of the fractional Brownian motion, they raise new issues when it comes to derivatives pricing. Using an original link between nearly unstable Hawkes processes and fractional volatility models, we compute the characteristic function of the log-price in rough Heston models. In the classical Heston model, the characteristic function is expressed in terms of the solution of a Riccati equation. Here we show that rough Heston models exhibit quite a similar structure, the Riccati equation being replaced by a fractional Riccati equation.
\end{abstract}


\section{Introduction}\label{intro}

The celebrated Heston model is a one-dimensional stochastic volatility model where the asset price $S$ follows the following dynamic: 
$$ dS_t = S_t \sqrt{V_t} dW_t $$
\begin{equation}
\label{heston}
dV_t = \lambda (\theta - V_t) dt + \lambda \nu \sqrt{V_t} dB_t.
\end{equation}
Here the parameters $\lambda$, $\theta$, $V_0$ and $\nu$ are positive, and $W$ and $B$ are two Brownian motions with correlation coefficient $\rho$, that is $\langle dW_t,dB_t\rangle=\rho dt$.\\

\noindent The popularity of this model is probably due to three main reasons:
\begin{itemize}
\item It reproduces well several important stylized facts of low frequency price data, namely leverage effect, time-varying volatility and fat tails, see \cite{bouchaud2003theory,christie1982stochastic,dragulescu2002probability,mandelbrot1997variation}.
\item It generates very reasonable shapes and dynamics for the implied volatility surface. Indeed, the ``volatility of volatility" parameter $\nu$ enables us to control the smile, the correlation parameter $\rho$ to deal with the skew, and the initial volatility $V_0$ to fix the at-the-money volatility level, see \cite{forde2012small,gatheral2011volatility,janek2011fx,poon2009heston}. Furthermore, as observed in markets and in contrast to local volatility models, in Heston model, the volatility smile moves in the same direction as the underlying and the forward smile does not flatten with time, see \cite{gatheral2011volatility,jacquier2013small,jacquier2016large,mazzon2015forward}.
\item There is an explicit formula for the characteristic function of the asset log-price, see \cite{heston1993closed}. From this formula, efficient numerical methods have been developed, allowing for instantaneous model calibration and pricing of derivatives, see \cite{albrecher2006little,carr1999option,kahl2005not,lewis2001simple}.
\end{itemize}

\noindent In the classical Heston model, the volatility follows a Brownian semi-martingale. However, it is shown in \cite{gatheral2014volatility} that for a very wide range of assets, historical volatility time-series exhibit a behavior which is much rougher than that of a Brownian motion. More precisely, dynamics of log-volatility are very well modeled by a fractional Brownian motion with Hurst parameter of order $0.1$. Furthermore, using a fractional Brownian motion with small Hurst index also enables us to reproduce very accurately the features of the volatility surface, see \cite{bayer2016pricing,gatheral2014volatility}. Finally, convincing microstructural foundations for rough volatility models are provided in \cite{eleuch2016micro,jaisson2015rough}, see also Section \ref{construction}.\\

\noindent Hence, in this paper, we are interested in the fractional versions of Heston model. Our main goal is to design an efficient pricing methodology for such models, in the spirit of the one introduced by Heston in the classical case. This is particularly important in fractional volatility models where the use of Monte-Carlo methods can be quite intricate due to the non-Markovian nature of the fractional Brownian motion, see \cite{bennedsen2015hybrid}.\\

\noindent We now define our so-called rough Heston model. Let us recall that a fractional Brownian motion $W^H$ with Hurst parameter $H\in(0,1)$ can be built through the Mandelbrot-van Ness representation:
\begin{equation}
\label{vanness}
W_t^H = \frac{1}{\Gamma(H+1/2)} \int_{-\infty}^0 \big((t-s)^{H-\frac{1}{2}} - (-s)^{H-\frac{1}{2}}\big)dW_s  + \frac{1}{\Gamma(H+1/2)} \int_0^t (t-s)^{H-\frac{1}{2}} dW_s.
\end{equation}
The kernel $(t-s)^{H-\frac{1}{2}}$ in \eqref{vanness} plays a central role in the rough dynamic of the fractional Brownian motion for $H<1/2$. In particular, one can show that the process
$$ \int_0^t (t-s)^{H-\frac{1}{2}} dW_s $$
has H\"older regularity $H-\varepsilon$ for any $\varepsilon > 0$. In order to allow for a rough behavior of the volatility in a Heston-type model, we naturally introduce the kernel $(t-s)^{\alpha -1}$ in a Heston-like stochastic volatility process as follows:
$$dS_t = S_t \sqrt{V_t} dW_t$$
\begin{equation}
\label{roughHeston}
 V_t = V_0 +\frac{1}{\Gamma(\alpha)} \int_0^t (t-s)^{\alpha - 1} \lambda (\theta - V_s)  ds  + \frac{1}{\Gamma(\alpha)} \int_0^t (t-s)^{\alpha - 1} \lambda \nu \sqrt{V_s}  dB_s.
\end{equation}
The parameters $\lambda$, $\theta$, $V_0$ and $\nu$ in \eqref{roughHeston} are positive and play the same role as in \eqref{heston}, and here also $W$ and $B$ are two Brownian motions with correlation $\rho$. The additional parameter $\alpha$ belongs to $(1/2,1)$ and governs the smoothness of the volatility sample paths.  More precisely, we show in this paper that the model is well-defined and that the volatility trajectories have almost surely H\"older regularity $\alpha - 1/2-\varepsilon$, for any $\varepsilon>0$. When $\alpha = 1$, Models \eqref{roughHeston} and \eqref{heston} coincide, and we retrieve the classical Heston model. Therefore it is natural to view \eqref{roughHeston} as a rough version of Heston model and to call it rough Heston model. Nevertheless, note that other definitions of rough Heston models can make sense, see \cite{guennoun2014asymptotic} for an alternative definition and some asymptotic results.\\

\noindent Our aim in this work is to derive a Heston-type formula for the characteristic function of the log-price in Model \eqref{roughHeston}. In the classical case ($\alpha=1$, Model \eqref{heston}), this formula is proved in \cite{heston1993closed}. It is obtained using the fact that Model \eqref{heston} is Markovian and time-homogeneous, and applying It\^o's formula to the function 
$$ L(t,a,V_t,S_t) = \mathbb{E}[e^{ia \log(S_T)}|{\cal{F}}_t],~~{\cal{F}}_t = \sigma(W_s,B_s; s \leq t),~~a \in \mathbb{R}.$$
The process $L$ being a martingale, the following Feynman-Kac partial differential equation for $L$ is easily obtained: 
$$-\partial_t L(t,a,S,V) =  \big{(}\lambda (\theta - V) \partial_v + \frac{1}{2} (\lambda \nu)^2 V \partial_{vv}^2+\frac{1}{2} S^2 V \partial_{ss}^2+\rho \nu \lambda S V \partial_{sv}^2 \big{)}L(t,a,S,V),$$ with boundary condition $L(T,a,S,V) = e^{ia \log(S)}$.
From this PDE, it can be checked that the characteristic function of the log-price $X_t= \log(S_t/S_0)$ satisfies
$$\mathbb{E}[e^{iaX_t}] = \exp\big{(}g(a,t)  + V_0 h(a,t) \big{)},$$
where $h$ is solution of the following Riccati equation:
\begin{equation}\label{classriccati} 
\partial_th = \frac{1}{2}(-a^2-ia)+ \lambda (i a \rho \nu-1)  h(a,s) + \frac{(\lambda \nu)^2}{2 } h^2(a,s),~~h(a,0) = 0,\end{equation}
and 
$$g(a,t) = \theta \lambda \int_0^t h(a,s) ds.$$
Solving this Riccati equation leads to the closed-form formula for the characteristic function of the log-price given in \cite{heston1993closed}.\\

\noindent In the case $\alpha<1$, the rough Heston model \eqref{roughHeston} is neither Markovian nor a semi-martingale. Hence the strategy initially used by Heston presented above seems very hard to adapt to our setting. Here we resort to a completely different and original approach based on point processes. Indeed, our methodology finds its root in the works \cite{eleuch2016micro,jaisson2015rough} which provide microstructural foundations to rough volatility models. In these papers, it is shown that some well-designed microstructure models, reproducing the stylized facts of modern financial markets at high frequency, give rise in the long run to rough volatility models. These microstructure models that we describe in more details in Section \ref{construction} are based on so-called nearly unstable Hawkes processes. In this paper, inspired by these results and using again Hawkes processes, we design a suitable sequence of point processes which converges to Model \eqref{roughHeston}. Exploiting the specific structure of our point processes, we derive their characteristic function, which leads us in the limit to that of the log-price in the rough Heston model \eqref{roughHeston}.\\ 

\noindent Our main result is that, quite surprisingly, the characteristic function of the log-price in rough Heston models exhibits the same structure as the one obtained in the classical Heston model. The difference is that the Riccati equation \eqref{classriccati} is replaced by a fractional Riccati equation, where a fractional derivative appears instead of a classical derivative. More precisely, we obtain
$$ \mathbb{E}[e^{iaX_t}] = \exp\big(g_1(a,t)  + V_0 g_2(a,t) \big),$$
where
$$ g_1(a,t) = \theta \lambda \int_0^t h(a,s) ds ,\quad g_2(a,t) = I^{1-\alpha}h(a,t),$$
and $h$ is a solution of the following fractional Riccati equation: 
$$ D^\alpha h = \frac{1}{2}(-a^2-ia)+ \lambda (i a \rho \nu-1)  h(a,s) + \frac{(\lambda \nu)^2}{2 } h^2(a,s),~~I^{1-\alpha}h(a,0) = 0,$$
with $D^\alpha$ and $I^{1-\alpha}$ the fractional derivative and integral operators defined in \eqref{defint} and \eqref{defdiff}. Remark that when $\alpha=1$, this result indeed coincides with the classical Heston's result. However, note that for $\alpha<1$, the solutions of such Riccati equations are no longer explicit. Nevertheless, they are easily solved numerically, see Section \ref{numerical}.\\

\noindent The paper is organized as follows. In Section \ref{construction}, we build a sequence of Hawkes-type processes which converges to the rough Heston model \eqref{roughHeston}. Then we study in Section \ref{CharacHawkes} the characteristic function of these processes and show in Section \ref{CharacHeston} that it enables us to derive the characteristic function of the log-price in Model \eqref{roughHeston}. One numerical illustration is given in Section \ref{numerical} and some proofs are relegated to Section \ref{proofs}. Finally, some useful technical results are given in an appendix.

\section{From Hawkes processes to rough Heston models}\label{construction}

We build in this section a sequence of Hawkes-type processes which converges to the rough Heston model \eqref{roughHeston}. This construction is inspired by the paper \cite{eleuch2016micro}. In this work, microstructural foundations for rough Heston models are provided. This is done designing suitable sequences of ultra high frequency price models which reproduce the stylized facts of modern markets microstructure and converge in the long run to rough Heston models. These microscopic price models are based on Hawkes processes. So that the reader can well understand the genesis of our original methodology to compute the characteristic function in rough Heston models, we recall here the main ideas and results in \cite{eleuch2016micro}.

\subsection{Microstructural foundations for rough Heston models}

\noindent In \cite{eleuch2016micro}, we consider a sequence of bi-dimensional Hawkes processes $(N^{T,+},N^{T,-})$ indexed by $T>0$ going to infinity\footnote{Of course by $T$ we implicitly mean $T_n$ with $n\in\mathbb N$ tending to infinity.} and with intensity
\begin{equation}
\label{hawkes}
\lambda_t^T = \begin{pmatrix} \lambda_t^{T,+}  \\ \lambda_t^{T,-}  \end{pmatrix} = \mu_T  \begin{pmatrix} 1 \\ 1 \end{pmatrix} + \int_0^t a_T \phi(t-s).  \begin{pmatrix} dN_s^{T,+}  \\ dN_s^{T,-}  \end{pmatrix},
\end{equation}
with 
$$ \phi = \begin{pmatrix} \varphi_1 & \varphi_3\\ \varphi_2 & \varphi_4  \end{pmatrix}.$$ Here the $\varphi_i$ are measurable non-negative deterministic functions and $\mu_T$ and $0<a_T<1$ are some deterministic sequences of positive real numbers, see \cite{bacry2013some} and the references therein for more details about the definition of Hawkes processes. Then in \cite{eleuch2016micro}, inspired by \cite{bacry2013modelling,bacry2013some,jaisson2015limit}, we consider the following ultra high frequency tick-by-tick model for the transaction price $P_t^T$: 
$$ P_t^T = N_t^{T,+} - N_t^{T,-}.$$ Hence $N_t^{T,+}$ represents the number of upward jumps of one tick of the transaction price over the period $[0,t]$ and  
$N_t^{T,-}$ the number of downward jumps. The relevance of this Hawkes-based modeling is that it enables us to encode very easily the most important stylized facts of high frequency markets in term of the parameters of the Hawkes process. We now give these stylized facts and their translation in term of the model parameters, referring to \cite{eleuch2016micro} for more details.
\begin{itemize}
\item Markets are highly endogenous: In the high frequency trading context, most orders have no real economic motivation. They are rather sent by algorithms as reaction to other orders. In the Hawkes framework, this amounts to work with so-called {\it nearly unstable Hawkes processes}. This means that the stability condition 
$$\mathcal{S}\big(\int_0^\infty a_T\phi(s)ds\big)<1,$$
where $\mathcal{S}$ denotes the spectral radius operator, should almost be saturated and that the intensity of exogenous orders, namely $\mu_T$, should be small, see \cite{eleuch2016micro,hardiman2013critical,jaisson2015rough,jaisson2015limit}. In term of model parameters, suitable constraints are therefore
$$a_T\rightarrow 1,~~\mathcal{S}\big(\int_0^\infty \phi(s)ds\big)=1,~~\mu_T\rightarrow 0.$$
\item It is not an easy task to make money with high frequency strategies on highly liquid electronic markets. Hence some ``no statistical arbitrage" mechanisms should be in force. 
We translate this assuming that in the long run, there are on average as many upward than downward jumps. This corresponds to the assumption
$$\varphi_1+\varphi_3=\varphi_2+\varphi_4.$$
\item Buying is not the same action as selling. This means that buy market orders and sell limit orders are not symmetric orders. To see this, consider for example a market maker, with an inventory which is typically positive. He is likely to raise the price by less following a buy order than to lower 
the price following the same size sell order. Indeed, its inventory becomes smaller after a buy order, which is a good thing for him, whereas it increases after a sell order. This creates a liquidity asymmetry on the bid and ask sides of the order book. This can be modeled in the Hawkes framework assuming that 
$$ \varphi_{3}=\beta \varphi_{2},$$
for some $ \beta > 1$. Hence, the matrix $\phi$ finally takes the form
$$ \phi = \begin{pmatrix} \varphi_1 & \beta \varphi_2  \\ \varphi_2 & \varphi_1 + (\beta-1) \varphi_2  \end{pmatrix}.$$
\item A significant amount of transactions is part of metaorders, which are large orders whose execution is split in time by trading algorithms. This is translated into a heavy tail assumption on the functions $\varphi_1$ and $\varphi_2$, namely that there exists $1/2<\alpha<1$ (typically around 0.6 in practice, see \cite{bacry2014estimation,hardiman2013critical}) and $C >0$ such that
$$\alpha x^{\alpha} \int_x^\infty  \varphi_1(s) + \beta \varphi_2(s) ds \underset{x \rightarrow \infty}{\rightarrow} C.$$
Furthermore, it is shown in \cite{jaisson2015rough} that for a given $\alpha$, there is only one way to make $\mu_T$ tends to zero and $a_T$ tends to one so that the limit of the price is not degenerate. More precisely,
$$(1-a_T) T^{\alpha} \underset{T \rightarrow \infty}{\rightarrow} \lambda^*,~~\mu_T T^{1-\alpha} \underset{T \rightarrow \infty}{\rightarrow} \mu,$$
for some positive $\lambda^*$ and $\mu$.
\end{itemize}
Under the above assumptions, it is proved in \cite{eleuch2016micro} that the properly rescaled microscopic price process
$$\sqrt{\frac{1-a_T}{\mu T^\alpha}} P_{tT}^T,~~t\in[0,1]$$
converges in law as $T$ tends to infinity to the following macroscopic price dynamic $P_t$:
$$
P_t = \frac{\sqrt{2}}{1 - \int_0^\infty (\varphi_1-\varphi_2)} \int_0^t \sigma_s dW_s,$$
\begin{equation}
\sigma_t^2 = \frac{1}{\Gamma(\alpha)} \int_0^t (t-s)^{\alpha-1} \lambda (1 - \sigma_s^2) ds + \frac{1}{\Gamma(\alpha)} \lambda \nu \int_0^t (t-s)^{\alpha-1} \sigma_s dB_s\label{sigma},
\end{equation}
where $(W,B)$ is a bi-dimensional correlated Brownian motion with correlation
$$ \rho =  \frac{1-\beta}{\sqrt{2(1+\beta^2)}}$$ 
and 
$$\nu = \sqrt{\frac{2(1+\beta^2)}{\lambda^* \mu (1+\beta)^2}},~~\lambda = \lambda^*\frac{\alpha}{C \Gamma(1-\alpha)}.$$
Hence this result shows that the main stylized facts of modern electronic markets naturally give rise to a very rough behavior of the volatility. Indeed, recall that the Hurst parameter corresponds to $\alpha-1/2$.\\ 

\noindent Inspired by this result, our idea is to study the characteristic function of some kind of microscopic price processes in order to deduce that of our rough Heston macroscopic price of interest \eqref{roughHeston}. However, the developments presented above cannot be directly applied and need to be adapted. Indeed, remark that in \eqref{sigma}, $\sigma_0 = 0$. This does not correspond to the case of \eqref{roughHeston}, where having a non-zero initial volatility is of course crucial for the model to be relevant in practice. Thus we need to modify the sequence of Hawkes-type processes to obtain a non-degenerate initial volatility in the limit. This is actually a non-trivial issue. However, this can be achieved replacing $\mu_T$ in \eqref{hawkes} by an inhomogeneous Poisson intensity $\hat{\mu}_T(t)$. We explain how such $\hat{\mu}_T(t)$ can be found in the next section.

\subsection{Finding the right Poisson rate}\label{Assumptions}
We work on a sequence of probability spaces $({\Omega}^T, {\cal{F}}^T,{\mathbb{P}}^T)$, indexed by $T>1$, on which $N^T = (N^{T,+},N^{T,-})$ is a bi-dimensional Hawkes process with intensity:
\begin{equation} 
\label{intensity}
\lambda_t^T=  \begin{pmatrix}\lambda_t^{T,+} \\  \lambda_t^{T,-}  \end{pmatrix} = \hat{\mu}_T(t) \begin{pmatrix} 1  \\ 1   \end{pmatrix}   + \int_0^t \phi^T(t-s). dN_s^T.
\end{equation}
For a given $T$, the probability space is equipped with the filtration $(\mathcal{F}_t^T)_{t\geq 0}$, where $\mathcal{F}_t^T$ is the $\sigma$-algebra generated by $(N_s^T)_{s\leq t}$. Since our goal is to design a sequence of processes leading in the limit to a rough Heston dynamic, we consider the same kind of assumptions on the matrix $\phi^T$ as those described in the previous section. However, here we can be very specific since we just need to find one convenient sequence of processes. That is why we make a particular choice for the heavy-tailed functions defining $\phi^T$, using Mittag-Leffler functions, see Section \ref{mittag} in Appendix for definition and some properties. Indeed, these functions are very convenient in order to carry out computations. More precisely, our assumptions on $\phi^T$ are as follows.
\begin{assumption}
\label{assumption1}
There exist $\beta\geq 0$, $1/2<\alpha<1$ and $\lambda>0$ such that 
$$a_T=1-\lambda T^{-\alpha},~~\phi^T = \varphi^T \chi,$$
where
$$\chi = \frac{1}{\beta +1}   \begin{pmatrix} 1 & \beta  \\ 1 & \beta   \end{pmatrix},~~\varphi^T = a_T \varphi,~~\varphi = f^{\alpha,1},$$
with $f^{\alpha,1}$ the Mittag-Leffler density function defined in Appendix.
\end{assumption}

\begin{rem}
\label{conv}
As in the previous section, we are working in the nearly unstable heavy tail case since 
$$ \int_0^{\infty} \varphi(s) ds = 1 $$ 
and $$ \alpha x^\alpha \int_x^\infty \varphi(t) dt  \underset{x \rightarrow \infty}{\longrightarrow} \frac{\alpha}{\Gamma(1-\alpha)}.$$
\end{rem}

\noindent We now give intuitions on how to find a suitable Poisson intensity $\hat{\mu}_T(t)$. The developments here are not very rigorous. They just aim at helping the reader to understand how our point processes sequence is designed. First, note that under Assumption \ref{assumption1},
$$ \lambda_t^{T,+} = \lambda_t^{T,-}.$$ The asymptotic behavior of the renormalized intensity processes $\lambda_t^{T,+}$ and $\lambda_t^{T,-}$ will give us that of the volatility in our limiting macroscopic price model. Thus, we need to understand the long term limit of $\lambda_t^{T,+}$. Let us write
$$M_t^T = (M_t^{T,+},M_t^{T,-}) = N_t^T - \int_0^t \lambda_s^T ds $$
for the martingale associated to the point process $N_t^T$. 
We easily obtain
$$\lambda_t^{T,+} = \hat{\mu}_T(t) + \int_0^t \varphi^T(t-s) \lambda_s^{T,+} ds + \frac{1}{1+\beta}\int_0^t \varphi^T(t-s) (dM_s^{T,+}+\beta dM_s^{T,-}).$$
Now let $$ \psi^T= \sum_{k\geq 1}  (\varphi^T)^{*k},$$
where  $(\varphi^T)^{*1} = \varphi^T $ and for $k>1$, $(\varphi^T)^{*k}(t) = \int_0^t \varphi^T(s)(\varphi^T)^{*(k-1)}(t-s)ds $. Using Lemma \ref{hopf} in Appendix together with Fubini theorem and the fact that $\psi^T*\varphi^T=\psi^T-\varphi^T$, we get
\begin{equation}
\label{lambda1}
\lambda_t^{T,+} = \hat{\mu}_T(t) + \int_0^t \psi^T(t-s) \hat{\mu}_T(s) ds + \frac{1}{1+\beta}\int_0^t \psi^T(t-s) (dM_s^{T,+}+\beta dM_s^{T,-}). 
\end{equation}
Following \cite{eleuch2016micro}, the inhomogeneous intensity $\hat{\mu}_T(t)$ should be of order $\mu_T$ with 
$$ \mu_T= \mu T^{\alpha-1},$$  where $\mu$ is some positive constant. 
In \cite{eleuch2016micro}, it is shown that the right normalization for the intensity in order to get a non-degenerate limit is to consider $(1-a_T)\lambda_{tT}^{T,+}/\mu_T$. The same applies here and thus we define the renormalized intensity
$$C_t^T= \frac{1-a_T}{\mu_T} \lambda_{tT}^{T,+}.$$ After obvious computations, this can be written
$$C_t^{T} = \frac{1-a_T}{\mu_T}\hat{\mu}_T(tT) + \int_0^t T(1-a_T) \psi^T\big(T(t-s)\big) \frac{\hat{\mu}_T(Ts)}{\mu_T} ds + \nu \int_0^t T (1-a_T) \psi^T\big(T(t-s)\big) \sqrt{C_s^T} dB_s^T,$$
where
$$ B_t^T = \int_0^{tT} \frac{dM_s^{T,+}+\beta dM_s^{T,-}}{\sqrt{T(\lambda_s^{T,+}+\beta^2 \lambda_s^{T,-})}},~~\nu =  \sqrt{\frac{1+\beta^2}{\lambda \mu (1+\beta)^2}}.$$
Using the fact that the Laplace transform $\hat{f}^{\alpha,\lambda}$ of the Mittag-Leffler density function $f^{\alpha,\lambda}$ is given by
$$ \hat{f}^{\alpha,\lambda}(z) = \frac{\lambda}{\lambda+z^\alpha},$$ we easily obtain that
\begin{equation}
\label{rho}
(1-a_T)T\psi^T(T.)=a_T f^{\alpha,\lambda},
\end{equation}
see Section \ref{mittag} in Appendix. This leads to the following expression for $C^T$: 
$$C_t^{T} = \frac{1-a_T}{\mu_T}\hat{\mu}_T(tT) + \int_0^t a_T f^{\alpha,\lambda}(t-s)\frac{\hat{\mu}_T(Ts)}{\mu_T} ds + \nu \int_0^t a_T f^{\alpha,\lambda}(t-s) \sqrt{C_s^T} dB_s^T.$$
Computing the quadratic variation of $B^T$, it is easy to see that it converges to a Brownian motion $B$. Now, if as in \cite{eleuch2016micro} we take
$\hat{\mu}_T(t) = \mu_T$, $C^T$ should then give in the limit a process $\sigma^2$ satisfying
$$ \sigma_t^2 =  F^{\alpha,\lambda}(t) + \nu \int_0^t f^{\alpha,\lambda}(t-s) \sigma_s dB_s,$$ where $$F^{\alpha,\lambda}(t) =\int_0^tf^{\alpha,\lambda}(u)du.$$ 
From Proposition \ref{uniqueness} in Section \ref{proofs}, this is equivalent to 
$$\sigma_t^2 = \frac{1}{\Gamma(\alpha)} \int_0^t (t-s)^{\alpha-1} \lambda (1 - \sigma_s^2) ds + \frac{1}{\Gamma(\alpha)} \lambda \nu \int_0^t (t-s)^{\alpha-1} \sigma_s dB_s,$$ which corresponds to \eqref{sigma}.
However, recall that we wish to obtain a limit where the initial volatility does not vanish, that is a process of the form
\begin{equation}
\label{sigmapos1}
\sigma_t^2 = \xi + \frac{1}{\Gamma(\alpha)} \int_0^t (t-s)^{\alpha-1} \lambda (1 - \sigma_s^2) ds + \frac{1}{\Gamma(\alpha)} \lambda \nu \int_0^t (t-s)^{\alpha-1} \sigma_s dB_s,
\end{equation}
with $\xi>0$. Again from Proposition \ref{uniqueness} in Section \ref{proofs}, the dynamic \eqref{sigmapos1} is equivalent to
\begin{equation*}
\sigma_t^2 = \xi\big(1-F^{\alpha,\lambda}(t)\big)+F^{\alpha,\lambda}(t) + \nu \int_0^t f^{\alpha,\lambda}(t-s) \sigma_s dB_s.
\end{equation*}
Using the same heuristic arguments as above, we see that we should obtain this dynamic in the limit provided we work with a process $C_t^T$ having the following expression: 
$$ C_t^T = (1-a_T) + \xi + (1-\xi)\int_0^t T(1-a_T)\psi\big(T(t-s)\big) ds + \nu \int_0^t T(1-a_T)\psi\big(T(t-s)\big) \sqrt{C_s^T}dB_s^T.$$
This is equivalent to
\begin{equation}
\label{lambda2} 
\lambda_t^{T,+} = \mu_T + \xi \mu_T \frac{1}{1-a_T} + \mu_T(1 - \xi ) \int_0^t \psi^T(t-s)ds+ \frac{1}{1+\beta}\int_0^t \psi^T(t-s) (dM_s^{T,+}+\beta dM_s^{T,-}). 
\end{equation}
Therefore, identifying parameters in \eqref{lambda1} and \eqref{lambda2}, this indicates that we should take $\hat{\mu}_T$ such that 
\begin{equation}
\label{eqaresoudre}  
\hat{\mu}_T(t) + \int_0^t \psi^T(t-s) \hat{\mu}_T(s) ds  =\mu_T +  \xi \mu_T \frac{1}{1-a_T} + \mu_T(1-\xi) \int_0^t \psi^T(t-s)ds.
\end{equation}
Using convolution by $\varphi^T$ together with the fact that $\psi^T*\varphi^T = \psi^T - \varphi^T$, we obtain from the left-hand side of \eqref{eqaresoudre}:
\begin{align*}
&\int_0^t \hat{\mu}_T(s) \varphi^T(t-s) ds + \int_0^t \int_0^s \psi^T(s-u) \hat{\mu}_T(u) du \varphi^T(t-s) ds   \\
&= \int_0^t \hat{\mu}_T(s) \varphi^T(t-s) ds + \int_0^t \int_0^{t-u} \psi^T(s) \varphi^T(t-u-s)  ds \hat{\mu}_T(u) du\\
&= \int_0^t \hat{\mu}_T(s) \varphi^T(t-s) ds + \int_0^t \big(\psi^T(t-u)-\varphi^T(t-u)\big)\hat{\mu}_T(u) du   \\
&= \int_0^t \psi^T(t-s) \hat{\mu}_T(s) ds.
\end{align*}
From the right-hand side of \eqref{eqaresoudre}, we get: 
\begin{align*}
&\int_0^t \varphi^T(t-s) (\mu_T +  \xi \mu_T \frac{1}{1-a_T}) ds + \mu_T(1-\xi) \int_0^t \varphi^T(t-s) \int_0^s  \psi^T(s-u) du ds \\
&= \mu_T(1 +  \xi \frac{1}{1-a_T}) \int_0^t \varphi^T(t-s)  ds + \mu_T(1-\xi) \int_0^t \int_0^{t-u}  \psi^T(s)\varphi^T(t-u-s)    ds du \\
&= \mu_T(1 +  \xi \frac{1}{1-a_T}) \int_0^t \varphi^T(t-s)  ds + \mu_T(1-\xi) \int_0^t  \big(\psi^T(t-u) - \varphi^T(t-u)\big) du.
\end{align*}
Consequently, the following equality should hold for a well-chosen $\hat{\mu}_T(s)$:
$$\int_0^t \psi^T(t-s) \hat{\mu}_T(s) ds  = \mu_T \xi ( \frac{1}{1-a_T}+1) \int_0^t \varphi^T(t-s)ds + \mu_T(1-\xi) \int_0^t \psi^T(t-s)ds.$$
This last equation together with \eqref{eqaresoudre} gives
\begin{equation}
\label{hatmu}
 \hat{\mu}_T(t) = \mu_T + \xi \mu_T \frac{1}{1-a_T} \big(1 - \int_0^t \varphi^T(t-s) ds\big) - \mu_T \xi\int_0^t \varphi^T(t-s)ds.
\end{equation}
Therefore, we should choose a non-homogenous baseline intensity $\hat{\mu}_T$ satisfying \eqref{hatmu}. In that case, we should recover the process \eqref{sigmapos1} for the limiting behavior of $C_t^T$.\\

\noindent In light of the preceding computations, we consider from now on the following assumption.

\begin{assumption}
\label{assumption2}
The baseline intensity $ \hat{\mu}_T$ is given by
$$\hat{\mu}_T(t) = \mu_T + \xi  \mu_T  \big( \frac{1}{1-a_T} ( 1- \int_0^t \varphi^T(s) ds)  - \int_0^t \varphi^T(s) ds \big),$$
with $\xi>0$ and $\mu_T = \mu T^{\alpha-1}$ for some $\mu>0$.
\end{assumption}

\begin{rem}
\label{mu}
Note that $\hat{\mu}_T$ can also be written as follows:
$$ \hat{\mu}_T(t) = \mu_T + \xi  \mu_T  \big(\frac{T^\alpha}{\lambda}  \int_t^\infty \varphi(s) ds  + \lambda T^{-\alpha} \int_0^t \varphi(s) ds \big).$$ This shows that $\hat{\mu}_T$ is a positive function and thus that the intensity process $\lambda_t^T$ in \eqref{intensity} is well-defined.
\end{rem}

\subsection{The rough limits of Hawkes processes}
We now give a rigorous statement about the limiting behavior of our specific sequence of bi-dimensional nearly unstable Hawkes processes with heavy tails.
For $t \in [0,1]$, we define
$$X_t^T= \frac{1-a_T}{T^\alpha \mu} N_{tT}^T,~~\Lambda_t^T= \frac{1-a_T}{T^\alpha \mu} \int_0^{tT} \lambda_{s}^T ds,~~Z_t^T = \sqrt{\frac{T^\alpha \mu}{1-a_T}} (X_t^T - \Lambda_t^T).$$
Using a similar approach as that in \cite{eleuch2016micro}, we obtain the following result whose proof is given in Section \ref{proofs}.
\begin{theorem}
\label{ConvergenceTheorem}
As $T \rightarrow \infty$, under Assumptions \ref{assumption1} and \ref{assumption2}, the process $\big{(} \Lambda_t^T, X_t^T, Z_t^T \big{)}_{t \in [0,1]}$ converges in law for the Skorokhod topology to $(\Lambda, X, Z)$ where
$$\Lambda_t = X_t = \int_0^t Y_s ds \begin{pmatrix} 1  \\ 1\end{pmatrix},~~Z_t = \int_0^t \sqrt{Y_s} \begin{pmatrix} dB_s^1  \\ dB_s^2   \end{pmatrix},$$ and $Y$ is the unique solution of the rough stochastic differential equation 
$$Y_t = \xi +  \frac{1}{\Gamma(\alpha)} \int_0^t (t-s)^{\alpha - 1} \lambda(1-Y_s) ds +\lambda \sqrt{\frac{1+\beta^2}{\lambda \mu(1+\beta^2)}} \frac{1}{\Gamma(\alpha)} \int_0^t (t-s)^{\alpha - 1} \sqrt{Y_s} dB_s,$$
where
$$ B = \frac{B^1 + \beta B^2}{\sqrt{1+\beta^2} }$$ 
and $(B^1,B^2) $ is a bi-dimensional Brownian motion. Furthermore, for any $\varepsilon>0$, $Y$ has H\"older regularity $\alpha - 1/2 - \varepsilon$.
\end{theorem}

\noindent Hence Theorem \ref{ConvergenceTheorem} shows that designing our sequence of bi-dimensional Hawkes processes in a suitable way, its limit is differentiable and its derivative exhibits a rough Cox-Ingersoll-Ross like behavior, with non-zero initial value. This is exactly what we need for the limiting volatility of our microscopic price processes. Indeed, thanks to Theorem \ref{ConvergenceTheorem}, we are now able to build such microscopic processes converging to the log-price in \eqref{roughHeston}. More precisely, for $\theta>0$, let us define 
\begin{equation}\label{microprice}
P^T= \sqrt{\frac{\theta}{2}} \sqrt{\frac{1-a_T}{T^\alpha \mu}}(N_{.T}^{T,+}- N_{.T}^{T,-}) -  \frac{\theta}{2} \frac{1-a_T}{T^\alpha \mu} N_{.T}^{T,+} = \sqrt{\frac{\theta}{2}}(Z^{T,+} - Z^{T,-}) - \frac{\theta}{2}X^{T,+}.\end{equation}
We have the following corollary of Theorem \ref{ConvergenceTheorem}.
\begin{corollary} 
\label{ConvergenceHeston}
As $T \rightarrow \infty$, under Assumptions \ref{assumption1} and \ref{assumption2}, the sequence of processes $ (P_{t}^T)_{t \in [0,1]}$ converges in law for the Skorokhod topology to 
$$ P_t = \int_0^t \sqrt{ V_s} dW_s  - \frac{1}{2} \int_0^t V_s ds, $$
where $V$ is the unique solution of the rough stochastic differential equation
$$ V_t = \theta \xi  + \frac{1}{\Gamma(\alpha)} \int_0^t (t-s)^{\alpha - 1} \lambda(\theta-V_s) ds + \lambda \sqrt{\frac{\theta (1+\beta^2)}{\lambda \mu(1+\beta)^2}} \frac{1}{\Gamma(\alpha)} \int_0^t (t-s)^{\alpha - 1} \sqrt{V_s} dB_s,$$
with $(W,B)$ a correlated bi-dimensional Brownian motion whose bracket satisfies
$$ d\langle W,B\rangle_t = \frac{1-\beta}{\sqrt{2(1+\beta^2)}} dt.$$
\end{corollary}

\noindent Thus, we have succeeded in building a sequence of microscopic processes $P^T$, defined by \eqref{microprice}, which converges to (the logarithm of) our rough Heston process of interest \eqref{roughHeston}. Now our goal is to use the result of Corollary \ref{ConvergenceHeston} to compute the characteristic function of the log-price in the rough Heston model \eqref{roughHeston}. This is done in the next two sections.

\section{The characteristic function of multivariate Hawkes processes}
\label{CharacHawkes}

We have seen in the previous section that our sequence of Hawkes-based microscopic price processes converges to the log-price in the rough Heston model \eqref{roughHeston}. Therefore, if we are able to compute the characteristic function for the microscopic price, its limit will give us that of the log-price in a rough Heston model. We actually provide a more general result here, deriving the characteristic function of a multivariate Hawkes process (recall that a bi-dimensional Hawkes process is the building block for our microscopic price process  \eqref{microprice}). Hence we extend here some results already proved in \cite{hawkes1974cluster} in the one-dimensional case.

\subsection{Cluster-based representation}\label{clusterHawkes}

To derive our characteristic function, the representation of Hawkes processes in term of clusters, see \cite{hawkes1974cluster}, is very useful. We recall it now.
Let us consider a $d$-dimensional Hawkes process $N = (N^{1},...,N^{d})$ with intensity
\begin{equation} 
\label{intensiteHawkes}
\lambda_t= \begin{pmatrix}\lambda_t^{1} \\ \vdots \\  \lambda_t^{d}  \end{pmatrix}= \mu(t)  + \int_0^t \phi(t-s). dN_s,
\end{equation}
where $\mu: \mathbb{R}_+ \rightarrow \mathbb{R}_+^d$ is locally integrable and $\phi: \mathbb{R}_+ \rightarrow \cal{M}^{\textbf{d}}(\mathbb{R}_+)$ has integrable components such that $$\mathcal{S}\big(\int_0^\infty \phi(s)ds\big)<1.$$
The law of such process can be described through a population approach. Consider that there are $d$ types of individuals and for a given type, an individual can be either a migrant or the descendant of a migrant. Then the dynamic goes as follows from time $t=0$: 
\begin{itemize}
\item Migrants of type $k \in \{1,..,d \}$ arrive as a non-homogenous Poisson process with rate $\mu_k(t)$.
\item Each migrant of type $k \in \{1,..,d \}$ gives birth to children of type $j \in \{1,..,d \}$ following a non-homogenous Poisson process with rate $\phi_{j,k}(t)$.
\item Each child of type $k \in \{1,..,d \}$ also gives birth to other children of type $j \in  \{1,..,d \}$ following a non-homogenous Poisson process with rate $\phi_{j,k}(t)$.
\end{itemize}
Then, for $k  \in \{1,..,d \}$, $N_t^k$ can be taken as the number up to time $t$ of migrants and children born with type $k$. Indeed, the population approach above and the theoretical characterization \eqref{intensiteHawkes} define the same point process law. 

\subsection{The result}

Let $L(a,t)$ be the characteristic function of the Hawkes process $N$:
$$ L(a,t) = \mathbb{E}[\exp(i a.N_t)],~~t \geq 0,~~a \in \mathbb{R}^d,$$ 
where $a.N_t$ stands for the scalar product of $a$ and $N_t$. The cluster-based representation of multivariate Hawkes processes enables us to show the following result, proved in Section \ref{demo}, for their characteristic function.
 
\begin{theorem} 
\label{thHawkes}
We have
$$ L(a,t)= \exp\big{(}\int_0^t \big(C(a,t-s)-\mathbf{1}\big).\mu(s) ds \big{)}, $$
where $C : \mathbb{R}^d \times \mathbb{R}_+ \rightarrow \mathbb{C}^d$ is solution of the following integral equation:
$$ C(a,t) = \exp\big{(} ia + \int_0^t \phi^*(s).(C(a,t-s)-\mathbf{1}) ds \big{)},$$ with $\phi^*(s)$ the transpose of $\phi(s)$.
\end{theorem}
\noindent From Theorem \ref{thHawkes}, we are able to derive in Section \ref{CharacHeston} the characteristic function of rough Heston models.

\subsection{Proof of Theorem \ref{thHawkes}}
\label{demo}
We now give the proof of Theorem \ref{thHawkes}, exploiting
 the population construction presented in Section \ref{clusterHawkes}. We start by defining $d$ auxiliary independent $d$-dimensional point processes $(\tilde{N}^{k,j})_{1 \leq j \leq d}$, $k \in \{ 1, ...,d \}$,  defined as follows for each given $k \in \{1,..,d \}$: 
\begin{itemize}
\item Migrants of type $j \in \{ 1, ...,d \}$ arrive as a non-homogenous Poisson process with rate $\phi_{j,k}(t)$.
\item Each migrant of type $j \in \{1,..,d \}$ gives birth to children of type $l \in \{1,..,d \}$ following a non-homogenous Poisson process with rate $\phi_{l,j}(t)$.
\item Each child of type $j \in \{1,..,d \}$ also gives birth to other children of type $l \in  \{1,..,d \}$ following a non-homogenous Poisson process with rate $\phi_{l,j}(t)$.
\end{itemize}
For a given $k  \in \{1,..,d \}$, $\tilde{N}_t^{k,j}$ corresponds to the number, up to time $t$, of migrants and children with type $j$. A simple but crucial remark is that  $(\tilde{N}^{k,j})_{1 \leq j \leq d}$ is actually also a multivariate Hawkes process with migrant rate $(\phi_{j,k})_{1 \leq j \leq d}$ and kernel matrix $\phi$. We write $L_k(a,t)$ for its characteristic function
$$ L_k(a,t)= \mathbb{E}\big[\exp(ia.(\tilde{N}_t^{k,j})_{1 \leq j \leq d})\big],~~t \geq 0,~~a \in \mathbb{R}^d.$$

\noindent Now let us come back to the initial Hawkes process of interest $N$ defined by \eqref{intensiteHawkes}. For each $k \in \{1,...,d\}$ and $t \geq 0$, let $N_t^{0,k}$ be the number of its migrants of type $k$ arrived up to time $t$. Recall that the $N^{0,k}$, $1 \leq k \leq d$, are independent Poisson processes with rates $\mu_k(t)$. We also define  $T_1^k<...<T_{N_t^{0,k}}^k \in [0,t]$ the arrival times of migrants of type $k$ of the Hawkes process $N$, up to time $t$. Using the population approach presented in Section \ref{clusterHawkes}, it is clear that at time $t$, the number of descendants of different types of a migrant of type $k$ arrived at time $T_u^k$ has the same law as $(\tilde{N}_{t-T_u^k}^{k,j})_{1 \leq j \leq d}$, where $\tilde{N}$ is taken independent from $N$. Consequently,  
\begin{equation}\label{decompH}
 N_t^k \underset{law}{=} N_t^{0,k} + \sum_{1 \leq j \leq d} \, \sum_{1 \leq l \leq N_t^{0,j}} \tilde{N}_{t-T_l^j }^{j,k,(l)},\end{equation}
where the $(\tilde{N}^{j,k,(l)})_{1 \leq k \leq d}$, $1 \leq j \leq d$, $l \in \mathbb{N}$ are independent copies of $(\tilde{N}^{j,k})_{1 \leq k \leq d}$, $1 \leq j \leq d$, also independent of $N^0= (N^{0,k})_{1\leq k \leq d}$.\\ 

\noindent From \eqref{decompH}, we derive that conditional on $N^0$,
\begin{align*}
\mathbb{E}\big[\exp(ia.N_t) | N^0\big] & = \exp(i a.N_t^0) \underset{1 \leq j \leq d}{\prod} \,\, \underset{1 \leq l \leq N_t^{0,j}}{\prod} \mathbb{E}\big[ \exp(ia.(\tilde{N}_{t-T_l^j }^{j,k,(l)})_{1 \leq k \leq d} | N^0\big{)}\big]\\
& = \exp(i a.N_t^0) \underset{1 \leq j \leq d}{\prod} \,\, \underset{1 \leq l \leq N_t^{0,j}}{\prod} L_j(a,t-T_l^j).
\end{align*}
Now, for a given $k \in \{1,...,d\}$, conditional on $N_t^{0,k}$, it is well-known that ($T_1^k,...,T_{N_t^{0,k}}^k$) has the same law as $(X_{(1)},...,X_{(N_t^{0,k})})$ the order statistics built from iid variables $(X_1,..,X_{N_t^{0,k}})$ with density $\displaystyle\frac{\mu_k(s) 1_{s\leq t}}{\int_0^t \mu_k(s)ds}$. Thus we get
$$ \mathbb{E}\big[ \exp(ia.N_t) | N_t^0\big]  = \exp(i a.N_t^0) \underset{1 \leq j \leq d}{\prod}  \big{(} \int_0^t L_j(a,t-s) \frac{\mu_j(s)}{\int_0^t \mu_j(s)ds} ds \big{)}^{N_t^{0,j}}.$$
Therefore,
$$
 L(a,t) = \underset{1 \leq j \leq d}{\prod} \exp\big{(} (\int_0^t e^{ia_j} L_j(a,t-s) \frac{\mu_j(s)}{\int_0^t \mu_j(s)ds}  ds -  1) {\int_0^t \mu_j(s)ds} \big{)}.$$
Thus we finally obtain 
\begin{equation}\label{lap1} 
L(a,t) = \exp\big(\sum_{1 \leq j \leq d} \int_0^t (e^{ia_j} L_j(a,t-s) - 1)\mu_j(s) ds\big).
\end{equation}
In the same way, since $(\tilde{N}^{k,j})_{1 \leq j \leq d}$ is a multivariate Hawkes process with migrant rate $(\phi_{j,k})_{1 \leq j \leq d}$ and kernel matrix $\phi$, we get
\begin{equation}\label{lap2}
L_k(a,t) = \exp\big(\sum_{1 \leq j \leq d} \int_0^t (e^{ia_j} L_j(a,t-s) - 1)\phi_{j,k}(s) ds\big).
\end{equation}
Let us define $$C(a,t) = \big(e^{ia_j}L_j(a,t)\big)_{1 \leq j \leq d}.$$
From \eqref{lap1}, we have that
$$ L(a,t) = \exp\big{(}  \int_0^t (C(a,t-s)-\mathbf{1}).\mu(s) ds \big{)}$$
and from \eqref{lap2}, we deduce that $C$ is solution of the following integral equation 
$$ C(a,t) = \exp\big{(} ia + \int_0^t \phi^*(s).(C(a,t-s)-\mathbf{1}) ds \big{)}.$$ 
This ends the proof of Theorem \ref{thHawkes}.

\section{The characteristic function of rough Heston models}
\label{CharacHeston}
We give in this section our main theorem, that is the characteristic function for the log-price in rough Heston models \eqref{roughHeston}. It is obtained combining the convergence result for Hawkes processes stated in Corollary \ref{ConvergenceHeston} together with the characteristic function for multivariate Hawkes processes derived in Theorem \ref{thHawkes}. We start with some intuitions about the result.
\subsection{Intuition about the result}
\label{idee}
We consider the rough Heston model \eqref{roughHeston}. The parameters of the dynamic in \eqref{roughHeston} are here given in term of those of the sequence of processes $P^T$ defined in \eqref{microprice}. More precisely, we set 
$$V_0 = \xi \theta,~~\rho = \frac{1-\beta}{\sqrt{2(1+\beta^2)}},~~\nu = \sqrt{\frac{\theta(1+\beta^2)}{\lambda \mu (1+\beta)^2}},$$ and $\lambda$ and $\theta$ are the same as those in the dynamic of $P^T$. Remark that this implies that $\rho \in (-1/\sqrt{2},1/\sqrt{2}]$.  
We also write $P_t = \log(S_t/S_0)$. From Corollary \ref{ConvergenceHeston}, we know that 
$$P^T= \sqrt{\frac{\lambda \theta}{2 \mu}} T^{-\alpha} (N_{.T}^{T,+}- N_{.T}^{T,-}) -  {\frac{\lambda \theta}{2 \mu}} T^{-2 \alpha} N_{.T}^{T,+}$$
converges in law to $P$ as $T$ tends to infinity, where $N^T = (N^{T,+},N^{T,-})$ is a sequence of bi-dimensional Hawkes processes satisfying Assumptions \ref{assumption1} and \ref{assumption2}. Let us write $L^T((a,b),t)$ for the characteristic function of the process $N^T$ at time $t$ at point $(a,b)$
and $L_p$ for the characteristic function of $P$. The convergence in law implies that of
$L^T((a_T^+,a_T^-),tT)$ towards $L_p(a,t),$ where 
$$a_T^+ =  a \sqrt{\frac{\lambda \theta}{2 \mu}} T^{-\alpha} -  a {\frac{\lambda \theta}{2 \mu}} T^{-2 \alpha},~~a_T^- = - a \sqrt{\frac{\lambda \theta}{2 \mu}} T^{-\alpha}.$$
From Theorem \ref{thHawkes}, we know that
$$ L^T\big((a_T^+,a_T^-),tT\big) = \exp \Big{(}\int_0^{tT}\hat{\mu}_T(s)\big((C^{T,+}((a_T^+,a_T^-),tT-s)-1) +(C^{T,-}((a_T^+,a_T^-),tT-s)-1)\big) ds\Big{)},$$
where $C^{T}((a_T^+,a_T^-),t) =\big{(}C^{T,+}((a_T^+,a_T^-),t),C^{T,-}((a_T^+,a_T^-),t) \big{)}\in \cal{M}^\mathbf{1\times 2}(\mathbb{C})$ is solution of 
$$  C^T\big((a_T^+,a_T^-),t\big) = \exp\big{(} i(a_T^+, a_T^-) +  \int_0^{t} \big(C^T((a_T^+,a_T^-),t-s) - (1,1)\big).\phi^T(s) ds \big{)}.$$
Now let $$Y^T(a,.)= \big(Y^{T,+}(a,.),Y^{T,-}(a,.)\big)= C^T\big((a_T^+,a_T^-),.T\big): [0,1]\rightarrow \cal{M}^\mathbf{1\times 2}(\mathbb{C}).$$
Using a change of variables, we easily get that $Y^T(a,.)$ is solution of the equation
\begin{equation}\label{decY} Y^T(a,t) = \exp\big{(} i(a_T^+, a_T^-) + T \int_0^t \big(Y^T(a,t-s) - (1,1)\big).\phi^T(Ts) ds \big{)}\end{equation}
and that
\begin{equation}\label{LTY}
 L^T(a_T^+,a_T^-,tT) = \exp \Big{(} \int_0^t \big(T^\alpha (Y^{T,+}(a,t-s) -1 ) + T^\alpha (Y^{T,-}(a,t-s) -1 )\big)\big(T^{1-\alpha} \hat{\mu}(sT) \big) ds \Big{)}.
 \end{equation}
Thanks to Remarks \ref{conv} and \ref{mu}, it is easy to see that
\begin{align*}
T^{1-\alpha} \hat{\mu}(sT) &= T^{1-\alpha} \mu_T + \xi T^{1-\alpha} \mu_T \big(\frac{T^\alpha}{\lambda} \int_{sT}^\infty \varphi(u) du + \lambda T^{-\alpha} \int_0^{sT}\varphi(u)du\big)\\
&= \mu \big(1+ \frac{\xi}{\lambda} s^{-\alpha} (sT)^\alpha \int_{sT}^\infty \varphi(u) du\big) + \mu \xi \lambda T^{-\alpha} \int_0^{sT}\varphi(u)du\\
&\underset{T \rightarrow \infty}{\longrightarrow} \mu \big(1+ \frac{\xi}{\lambda \Gamma(1-\alpha)} s^{-\alpha}\big).
\end{align*}
Now, the convergence of  $L^T(a_T^+,a_T^-,tT) $ as $T$ goes to infinity implies that of 
$$\int_0^t \big{(} T^\alpha (Y^{T,+}(a,t-s) -1 ) + T^\alpha (Y^{T,-}(a,t-s) -1 )  \big{)}  \big(T^{1-\alpha} \hat{\mu}(sT) \big) ds.$$ Thus, we can expect that as $T$ goes to infinity, $T^{\alpha}(Y^T(a,t) - (1,1))$ converges to some function $(c(a,t),d(a,t))$ (recall that the developments in this section are not rigourous and just aim at giving intuitions for the main theorem). Furthermore, using that $(Y^T(a,t) - (1,1))=\mathcal{O}(T^{-\alpha})$ together with \eqref{LTY}, we get 
\begin{align*}
Y^T(a,t) - (1,1) &= \log\big(Y^T(a,t)\big) + \frac{1}{2} \big(Y^T(a,t) - (1,1)\big)^2 + o(T^{-2 \alpha})(t)\\
&= i a \sqrt{\frac{\lambda \theta}{2 \mu}} (1,-1)T^{-\alpha} - i a {\frac{\lambda \theta}{2 \mu}} (1,0) T^{-2 \alpha} +  T \int_0^t \big(Y^T(a,t-s) - (1,1)\big).\phi^T(Ts) ds\\
& + \frac{1}{2} \big(Y^T(a,t) - (1,1)\big)^2 + o(T^{-2 \alpha})(t),
\end{align*}
where the logarithm\footnote{The complex logarithm is defined on $\mathbb C/\mathbb R^-$ by $\log(z) = \log(|z|) + i \arg(z)$, with $\arg(z) \in (-\pi,\pi]$.}  function is applied on each component of $Y^T(a,t)$. Then, remarking that $\chi^2 = \chi$ and using a change of variables, we get
$$\sum_{k \geq 1}{\big(T \phi^T(T.)\big)^{*k}}= \sum_{k \geq 1}{T (\phi^T)^{*k}(T.)}= T \sum_{k \geq 1} (\varphi^T)^{*k}(T.) \chi^k= T \psi^T(T.) \chi.$$
From \eqref{rho}, we finally deduce that
\begin{equation*}
\sum_{k \geq 1}{\big(T \phi^T(T.)\big)^{*k}} = a_T \frac{T^\alpha}{\lambda} f^{\alpha,\lambda} \chi =a_T  \frac{T^{\alpha}}{\lambda(\beta + 1)}   f^{\alpha,\lambda} \begin{pmatrix} 1 & \beta   \\ 1 &   \beta  \end{pmatrix}.
\end{equation*}
Since $(1,-1).\chi = 0$, using Lemma \ref{hopf} in Appendix we obtain
\begin{align*}
Y^T(a,t) - (1,1) &= i a \sqrt{\frac{\lambda \theta}{2 \mu}} (1,-1)T^{-\alpha} - i a  {\frac{ a_T \theta}{2 \mu (\beta +1) }} F^{\alpha,\lambda}(t) (1,\beta) T^{- \alpha} \\
&+  \frac{a_TT^{\alpha}}{2\lambda(\beta + 1)} \int_0^t f^{\alpha,\lambda}(s)\big(Y^T(a,t-s) - (1,1)\big)^2. \begin{pmatrix} 1 & \beta   \\ 1 &   \beta  \end{pmatrix} ds  + o(T^{-\alpha})(t).
\end{align*}
Therefore, we can expect that 
 $$ c(a,t) = i a \sqrt{\frac{\lambda \theta}{2 \mu}} - i a  {\frac{ \theta}{2 \mu (\beta +1) }} F^{\alpha,\lambda}(t) +  \frac{1}{2\lambda(\beta + 1)} \int_0^t f^{\alpha,\lambda}(s)  \big(c^2(a,t-s)+d^2(a,t-s) \big)ds,$$
 $$ d(a,t) = - i a \sqrt{\frac{\lambda \theta}{2 \mu}} - i a  {\frac{\beta \theta}{2 \mu (\beta +1) }} F^{\alpha,\lambda}(t) +  \frac{\beta}{2\lambda(\beta + 1)} \int_0^t f^{\alpha,\lambda}(s)\big(c^2(a,t-s)+d^2(a,t-s)\big)ds$$
and 
$$ L_p(a,t) = \exp\big{(} \int_0^t g(a,s) ds + \frac{V_0}{\theta \lambda} \frac{1}{\Gamma(1-\alpha)} \int_0^t g(a,s) (t-s)^{-\alpha} ds \big{)},$$
with $g = \mu ( c+d)$. We give in the next section a rigorous statement for this result.
\subsection{Main result}
We define the fractional integral of order $r \in (0,1]$ of a function $f$ as
\begin{equation} 
\label{defint}
I^r f(t) = \frac{1}{\Gamma(r)} \int_0^t (t-s)^{r-1} f(s) ds,  
\end{equation}
whenever the integral exists, and the fractional derivative of order $r \in [0,1)$ as
\begin{equation} 
\label{defdiff}
D^r f(t) = \frac{1}{\Gamma(1-r)} \frac{d}{dt} \int_0^t (t-s)^{-r} f(s) ds,  
\end{equation}
whenever it exists.
The following theorem, proved in Section \ref{proofs}, is the main result of the paper.
\begin{theorem} \label{CharacRough}
Consider the rough Heston model \eqref{roughHeston} with a correlation between the two Brownian motions $\rho$ satisfying $\rho \in (-1/\sqrt{2},1/\sqrt{2}]$. For all $t\geq 0$, we have 
\begin{equation}\label{defL_p}L_p(a,t) = \exp\big{(} \theta \lambda I^1 h(a,t)  + V_0 I^{1-\alpha}h(a,t)\big{)},\end{equation}
where $h$ is solution of the fractional Riccati equation
\begin{equation}
\label{FracEDP}
D^\alpha h(a,t) = \frac{1}{2}(-a^2-ia)+ \lambda (i a \rho \nu-1)  h(a,s) + \frac{(\lambda \nu)^2}{2 } h^2(a,s),~~I^{1-\alpha}h(a,0) = 0,
\end{equation}
which admits a unique continuous solution.
\end{theorem}
\noindent Thus we have been able to obtain a semi-closed formula for the characteristic function in rough Heston models. This means that pricing of European options becomes an easy task in this model, see Section \ref{numerical}. For $\alpha=1$, we retrieve the classical Heston formula. For $\alpha<1$, the formula is almost the same. The difference is essentially only in that in the Riccati equation, the classical derivative is replaced by a fractional derivative. The drawback is that such fractional Riccati equations do not have explicit solutions. However, they can be solved numerically almost instantaneously, see Section \ref{numerical}. Finally, note that this strong link between Hawkes processes and (rough) Heston models is probably natural since both of them exhibit some kind of affine structure (although infinite-dimensional).

\section{Numerical application}\label{numerical}

\subsection{Numerical scheme}
We explain in this section how to compute numerically the characteristic function of the log-price in a rough Heston model. By Theorem \ref{CharacRough}, $L_p(a,t)$ is entirely defined through the fractional Riccati equation \eqref{FracEDP} 
$$ D^\alpha h(a,t) = F\big(a,h(a,t)\big),~~I^{1-\alpha}h(a,0) = 0,$$
where 
$$ F(a,x) =  \frac{1}{2}(-a^2-ia)+ \lambda (i a \rho \nu-1) x + \frac{(\lambda \nu)^2}{2 } x^2.$$
Several schemes for solving numerically \eqref{FracEDP} can be found in the literature. Most of them are based on the idea that \eqref{FracEDP} implies the following Volterra equation: 
\begin{equation}
\label{Volterra}
h(a,t) = \frac{1}{\Gamma(\alpha)} \int_0^t (t-s)^{\alpha-1} F\big(a,h(a,s)\big) ds.
\end{equation}
Then one develops numerical schemes for \eqref{Volterra}. Here we choose the well-known fractional Adams method investigated in \cite{diethelm2002predictor,diethelm2004detailed,diethelm1998fracpece}. The idea goes as follows. Let us write $g(a,t) = F\big(a,h(a,t)\big)$. Over a regular discrete time-grid $(t_k)_{k\in \mathbb{N}}$ with mesh $\Delta$ $(t_k=k\Delta)$, we estimate
$$h(a,t_{k+1}) = \frac{1}{\Gamma(\alpha)} \int_0^{t_{k+1}} (t_{k+1}-s)^{\alpha-1} g(a,s) ds $$
by
$$ \frac{1}{\Gamma(\alpha)} \int_0^{t_{k+1}} (t_{k+1}-s)^{\alpha-1} \hat{g}(a,s) ds,$$
where 
$$ \hat{g}(a,t) = \frac{t_{j+1}-t}{t_{j+1}-t_j} \hat{g}(a,t_{j})+ \frac{t-t_j}{t_{j+1}-t_j} \hat{g}(a,t_{j+1}),~~t\in [t_j,t_{j+1}),~~0 \leq j \leq k.$$
This corresponds to a trapezoidal discretization of the fractional integral and leads to the following scheme: 
\begin{equation}
\label{implicit}
\hat{h}(a,t_{k+1}) =  \sum_{0\leq j \leq k} a_{j,k+1} F\big(a,\hat{h}(a,t_j)\big)+ a_{k+1,k+1} F\big(a,\hat{h}(a,t_{k+1})\big), 
\end{equation}
with 
$$ a_{0,k+1} = \frac{\Delta^\alpha}{\Gamma(\alpha+2)} \big(k^{\alpha+1} - (k-\alpha)(k+1)^\alpha \big),$$
\begin{equation} 
\label{suiteA}
a_{j,k+1} = \frac{\Delta^\alpha}{\Gamma(\alpha+2)} \big((k-j+2)^{\alpha+1}+ (k-j)^{\alpha+1} - 2 (k-j+1)^{\alpha+1}\big),~~1 \leq j \leq k,
\end{equation}
and
$$ a_{k+1,k+1} = \frac{\Delta^\alpha}{\Gamma(\alpha+2)}.$$
However, $\hat{h}(a,t_{k+1})$ being on both sides of \eqref{implicit}, this scheme is implicit. Thus, in a first step, we compute a pre-estimation of  $\hat{h}(a,t_{k+1})$ based on a Riemann sum that we then plug into the trapezoidal quadrature. This pre-estimation, called predictor and that we denote by $\hat{h}^P(a,t_{k+1})$ is defined by 
$$\hat{h}^P(a,t_{k+1})=  \frac{1}{\Gamma(\alpha)} \int_0^t (t-s)^{\alpha-1} \tilde{g}(a,s) ds,$$
with
$$\tilde{g}(a,t) = \hat{g}(a,t_j),~~t\in [t_j,t_{j+1}),~~0 \leq j \leq k.$$
Therefore,
\begin{equation*}
\hat{h}^P(a,t_{k+1}) =  \sum_{0\leq j \leq k} b_{j,k+1} F\big(a,\hat{h}(a,t_j)\big), 
\end{equation*}
where
\begin{equation*}
b_{j,k+1} = \frac{\Delta^\alpha}{\Gamma(\alpha+1)}\big((k-j+1)^\alpha - (k-j)^\alpha\big),~~ 0 \leq j \leq k.
\end{equation*}
Thus, the final explicit numerical scheme is given by
\begin{equation*}
\hat{h}(a,t_{k+1}) =  \sum_{0\leq j \leq k} a_{j,k+1} F\big(a,\hat{h}(a,t_j)\big)  + a_{k+1,k+1} F\big(a,\hat{h}^P(a,t_j)\big),~~\hat{h}(a,0)=0, 
\end{equation*}
where the weights $a_{j,k+1}$ are defined in \eqref{suiteA}. Theoretical guarantees for the convergence of this scheme are provided in \cite{li2009fractional}. In particular, it is shown that for given $t>0$ and $a \in \mathbb{R}$,
$$ \underset{t_j \in [0,t]}{\max} |\hat{h}(a,t_j)- {h}(a,t_j)| = o(\Delta)$$
and
$$ \underset{t_j \in [\varepsilon,t]}{\max} |\hat{h}(a,t_j)- {h}(a,t_j)| = o(\Delta^{2-\alpha}),$$
for any $\varepsilon > 0$.

\subsection{One numerical illustration}
We consider the rough Heston model \eqref{roughHeston} with the following parameters:
$$\lambda = 2,~~\rho = -0.5,~~V_0= 0.4,~~\nu = 0.05,~~\theta = 0.04.$$
To compute $L_p(a,t)$, we use the numerical scheme presented above to solve Riccati equation and then plug the numerical solution into \eqref{defL_p}. 
Once the characteristic function is obtained, classical methods are available to obtain call prices
$$C(K,T) = \mathbb{E}[S_T - K]_+,$$
see \cite{carr1999option,itkin2005pricing,lewis2001simple} and the survey \cite{schmelzle2010option}. In our case, we use Lewis method, see \cite{lewis2001simple}. Here we display the term structure of the at-the-money skew, that is the derivative of the implied volatility with respect to log-strike for at-the-money calls. We compute it for $\alpha=1$ (classical Heston) and $\alpha=0.6$ (rough Heston with Hurst parameter equal to 0.1).  
\begin{center}
\includegraphics[scale=0.65]{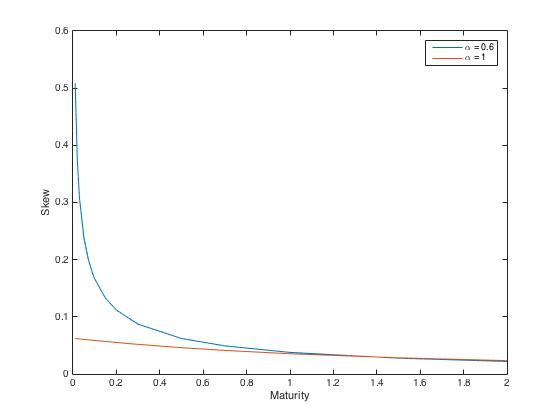}
\captionof{figure}{At-the-money skew as a function of maturity for $\alpha = 1 \text{ and } \alpha =0.6$}
\label{fig1}
\end{center}
We see that in the rough case, the skew explodes when maturity goes to zero, whereas it remains flat in the classical Heston case. This is a remarkable feature of rough-volatility models, very important for practical applications, see \cite{bayer2016pricing,fukasawa2011asymptotic,jaisson2015rough}.

\section{Proofs}
\label{proofs}
In the sequel, $c$ denotes a constant that may vary from line to line.
\subsection{Proof of Theorem \ref{ConvergenceTheorem}}
The proof of Theorem \ref{ConvergenceTheorem} is close to the one given in \cite{eleuch2016micro} for the convergence of a microscopic price model to a Heston-like dynamic. The main difference is that we have to deal here with a time-varying baseline intensity $\hat{\mu}_T$, which we have introduced to get a non-zero initial volatility in the limit. As in \cite{eleuch2016micro}, we start by showing the C-tightness of $(\Lambda^T,X^T,Z^T)$.
\subsubsection{C-tightness of $(\Lambda^T,X^T,Z^T)$}
We have the following proposition.
\begin{prop}\label{prop1}
Under Assumptions \ref{assumption1} and  \ref{assumption2}, the sequence $(\Lambda^T,X^T,Z^T)$ is C-tight and
$$ \underset{t \in [0,1]}{\sup} \|\Lambda_t^T-X_t^T\| \underset{T \rightarrow \infty}{\longrightarrow} 0$$
in probability. Moreover, if $(X,Z)$ is a possible limit point of $(X^T,Z^T)$, then $Z$ is a continuous martingale with $[Z,Z] = diag(X)$.
\end{prop}
\noindent\textsc{Proof}:
\paragraph{C-tightness of $X^T$ and $\Lambda^T$} Recall that as in \eqref{lambda1}, we can write
$$ \lambda^{T,+}_t =  \lambda^{T,-}_t = \hat{\mu}_T(t) + \int_0^t \psi^T(t-s)  \hat{\mu}_T(s) ds + \frac{1}{\beta+1}  \int_0^t \psi^T(t-s) (dM_s^{T,+}+\beta dM_s^{T,-}),$$where
$$ M_t^T= (M_t^{T,+},M_t^{T,-}) =N_t^T - \int_0^t \lambda_s^T ds  $$
is a martingale. Using that $\displaystyle\int_0^. (f*g) = (\int_0^. f)*g$, we get
$$ \mathbb{E}[N_{ T}^{T,+}] =  \mathbb{E}[N_{ T}^{T,-}] = \mathbb{E}\big[\int_0^{T}\lambda_s^{T,+}ds\big] = \int_0^{ T} \hat{\mu}_T(s) ds +  \int_0^{ T} \psi^T( T-s) \big(\int_0^s  \hat{\mu}_T(u) du \big) ds.$$
Consequently, $\hat{\mu}$ being a positive function and using that
$$ 1+ \int_0^\infty \psi^T(s) ds =1 + \sum_{k \geq 1} \int_0^\infty (\varphi^T)^{*k} = \sum_{k \geq 0} (a_T)^k = \frac{T^\alpha}{\lambda},$$
we obtain
$$\mathbb{E}[N_{ T}^{T,+}] \leq \int_0^{T} \hat{\mu}_T(s) ds \big(1+ \int_0^\infty \psi^T(s) ds\big)\leq \frac{1}{\lambda} T^{\alpha+1} \int_0^{1} \hat{\mu}_T(Ts) ds.$$
Moreover, from the definition of $\hat{\mu}$ and Remark \ref{conv}, we have 
$$
\int_0^{1} \hat{\mu}_T(Ts) ds = \mu T^{\alpha-1} \big{(} 1 + \xi \int_0^{1} s^{-\alpha} \frac{(sT)^\alpha}{\lambda} \int_{sT}^\infty \varphi(u) du ds + \lambda T^{-\alpha} \int_0^{1} \int_0^{sT} \varphi(u) du ds \big{)} 
\leq c T^{\alpha-1}.$$
Hence  $\mathbb{E}[N_{ T}^{T,+}] \leq  c T^{2 \alpha}$ and therefore
$$\mathbb{E}[X_{1}^T] = \mathbb{E}[\Lambda_{1}^T]\leq c,$$ for each component.
Each component of $X^T$ and $\Lambda^T$ being increasing, we deduce the tightness of each component of $(X^T,\Lambda^T)$. Furthermore, the maximum jump size of $X^T$ and $\Lambda^T$ being $\frac{1-a_T}{T^\alpha \mu}$ which goes to zero, the C-tightness of $(X^T,\Lambda^T)$ is obtained from Prop.VI-3.26 in \cite{jacod2013limit}.
\paragraph{C-tightness of $Z^T$}
It is easy to check that
$$ \langle Z^T,Z^T\rangle = diag(\Lambda^T),$$ which is C-tight. From Theorem VI-4.13 in \cite{jacod2013limit}, this gives the tightness of $Z^T$. The maximum jump size of $Z^T$ vanishing as $T$ goes to infinity, we obtain that $Z^T$ is C-tight.
\paragraph{Convergence of $X^T-\Lambda^T$}
We have
$$X_t^T-\Lambda_t^T = \frac{1-a_T}{T^\alpha \mu} M_{tT}^T.$$ 
From Doob's inequality, we get that for each component
$$\mathbb{E}\big[\underset{t \in [0,1]}{\sup}|\Lambda_t^T-X_t^T|^2\big]\leq c T^{-4 \alpha} \mathbb{E}[M_{ T}^T]^2.
$$
Since $[M^T,M^T]=N^T$, we deduce
$$\mathbb{E}\big[\underset{t \in [0,1]}{\sup}|\Lambda_t^T-X_t^T|^2\big]\leq c T^{-4 \alpha} \mathbb{E}[N_{ T}^T]\leq c T^{-2 \alpha}.
$$
This gives the uniform convergence to zero in probability of $X^T-\Lambda^T$.

\paragraph{Limit of $Z^T$}
Let $(X,Z)$ be a limit point of $(X^T,Z^T)$. We know that $(X,Z)$ is continuous and from Corollary IX-1.19 in \cite{jacod2013limit}, $Z$ is a local martingale. Moreover, since
$$ [Z^T,Z^T] = diag(X^T),$$
using Theorem VI-6.26 in \cite{jacod2013limit}, we get that $[Z,Z]$ is the limit of  $[Z^T,Z^T]$ and $[Z,Z]=diag(X)$. By Fatou's lemma, the expectation of $[Z,Z]$ is finite and therefore $Z$ is a martingale.
\qed

\subsubsection{Convergence of $X^{T}$ and $Z^{T}$}
First remark that since $$ \underset{t \in [0,1]}{\sup} |\Lambda_t^T-X_t^T| \underset{T \rightarrow \infty}{\longrightarrow} 0$$
and 
$$\Lambda_t^{T,+} = \Lambda_t^{T,-},$$
we get
$$\underset{t \in [0,1]}{\sup} |X_t^{T,+}-X_t^{T,-}| \underset{T \rightarrow \infty}{\longrightarrow} 0.$$
Therefore, if a subsequence of $X_t^{T,+}$ converges to some $X$, then the associated subsequence of $X_t^{T,-}$ converges to the same $X$.
We have the following proposition for the limit points of  $X_t^{T,+}$ and  $X_t^{T,-}$. 
\begin{prop}
\label{ConvergenceX}
\noindent If $(X,X,Z^+,Z^-)$ is a possible limit point for $(X^{T,+},X^{T,-},Z^{T,+},Z^{T,-})$, then $(X_t,Z_t^+,Z_t^-)$ can be written
$$X_t = \int_0^t Y_s ds,~~Z_t^+ = \int_0^t \sqrt{Y_s} dB_s^1,~~Z_t^- = \int_0^t \sqrt{Y_s} dB_s^2,$$ where $(B_1,B_2)$ is a bi-dimensional Brownian motion and $Y$ is solution of
\begin{equation}\label{defSDEY}
Y_t =  \xi\big(1-F^{\alpha,\lambda}(t)\big) + F^{\alpha,\lambda}(t) + \sqrt{ \frac{1+\beta^2}{\lambda \mu (1+\beta)^2}}  \int_0^t f^{\alpha,\lambda}(t-s) \sqrt{Y_s} dB_s,\end{equation}
with 
$$ B = \frac{B^1 + \beta B^2}{\sqrt{1+\beta^2}}. $$
Furthermore, for any  $\varepsilon>0$, $Y$ has H\"older regularity $\alpha - 1/2 - \varepsilon$.
\end{prop} 
\noindent \textsc{Proof}:\\

\noindent First recall that $\lambda_t^{T,+} = \lambda_t^{T,-}$ and note that using similar computations as in Section \ref{Assumptions}, we can write
$$\lambda_t^{T,+} =\mu_T + \mu_T \int_0^t  \psi^T(t-s)  ds + \xi \mu_T \big(\frac{1}{1-a_T} - \int_0^t \psi^T(t-s) ds\big)\\ 
+ \frac{1}{\beta+1}\int_0^t  \psi^T(t-s)  (dM_s^{T,+} + \beta dM_s^{T,-}).$$
Then using Fubini theorem together with the fact that $\displaystyle\int_0^. (f*g) = (\int_0^. f)*g$, we get
\begin{align*}
\int_0^t \lambda_s^{T,+} ds &=  \mu_T t  + \mu_T \int_0^t  \psi^T(t-s) s  ds + \xi \mu_T \big(\frac{t}{1-a_T} - \int_0^t \psi^T(t-s) s ds\big) \\
&+ \frac{1}{\beta+1}\int_0^t  \psi^T(t-s)  (M_s^{T,+} + \beta M_s^{T,-}) ds. 
\end{align*}
Therefore, for $t \in [0,1]$, we have the decomposition
\begin{equation} 
\label{aux7}
\Lambda_t^{T,+} = \Lambda_t^{T,-} =T_1 + T_2 + T_3, 
\end{equation}
with
$$ T_1 = (1-a_T) t,$$
$$ T_2 = T (1-a_T)  \int_0^t  \psi^T\big(T(t-s)\big) s  ds + \xi  \big{(}t - T (1-a_T)  \int_0^t  \psi^T\big(T(t-s)\big) s  ds \big{)}, $$
$$ T_3 = \frac{1}{\sqrt{\lambda \mu (1+\beta)^2}}  \int_0^t  T (1-a_T) \psi^T\big(T(t-s)\big) (Z_s^{T,+} + \beta Z_s^{T,-}) ds.$$
Now recall that we have shown in \eqref{rho} that 
$$ T(1-a_T) \psi(T.) = a_T f^{\alpha,\lambda}.$$
Thus
$$ T_2 \underset{T \rightarrow \infty}{\longrightarrow}   \int_0^t f^{\alpha,\lambda}(t-s)  s  ds + \xi  \big{(}t - \int_0^t f^{\alpha,\lambda}(t-s)  s  ds \big{)}$$ and
$$ T_3 \underset{T \rightarrow \infty}{\longrightarrow}  \frac{1}{\sqrt{\lambda \mu (1+\beta)^2}}  \int_0^t f^{\alpha,\lambda}(t-s)  (Z_s^+ + \beta Z_s^-) ds.$$
Therefore, letting $T$ go to infinity in \eqref{aux7}, we obtain using Proposition \ref{prop1} that $X$ satisfies
$$ X_t =  \int_0^t f^{\alpha,\lambda}(t-s)  s  ds + \xi  \big{(}t - \int_0^t f^{\alpha,\lambda}(t-s)  s  ds \big{)} + \frac{1}{\sqrt{\lambda \mu (1+\beta)^2}}  \int_0^t f^{\alpha,\lambda}(t-s)  (Z_s^+ + \beta Z_s^-) ds.$$
In the same way as for the proof of Theorem 3.2 in \cite{jaisson2015rough}, we show that
$$ X_t = \int_0^t Y_s ds,$$
where $Y$ satisfies
$$ Y_t =  F^{\alpha,\lambda}(t) + \xi \big(1-F^{\alpha,\lambda}(t)\big)+  \frac{1}{\sqrt{\lambda \mu (1+\beta)^2}}  \int_0^t f^{\alpha,\lambda}(t-s) (dZ_s^+ + \beta dZ_s^-).$$
Since, by Proposition \ref{prop1},
$$[Z,Z]=  \int_0^t Y_s ds \begin{pmatrix} 1 & 0\\ 0 & 1  \end{pmatrix},$$ 
we can apply Theorem V-3.9 in \cite{revuz1999continuous} to show the existence of a bi-dimensional Brownian motion $(B^1,B^2)$ such that
$$ Z_t^+ =  \int_0^t \sqrt{Y_s} dB_s^1,~~Z_t^- =  \int_0^t \sqrt{Y_s} dB_s^2.$$
Finally, we define the following Brownian motion:
$$ B = \frac{B^1 + \beta B^2}{\sqrt{1+\beta^2}}.$$
Then, in the same way as for the proof of Theorem 3.2 in \cite{jaisson2015rough}, we get that $Y$ satisfies
$$ Y_t =  F^{\alpha,\lambda}(t) + \xi \big(1-F^{\alpha,\lambda}(t)\big)+  \sqrt{\frac{1+\beta^2}{\lambda \mu (1+\beta)^2}}  \int_0^t f^{\alpha,\lambda}(t-s) \sqrt{Y_s} dB_s,$$
and has H\"older regularity $\alpha-1/2-\varepsilon$ for any $\varepsilon>0$. 
\qed

\subsubsection{End of the proof of Theorem \ref{ConvergenceTheorem}}
We now recall the following proposition stating that the process $Y$ is uniquely defined by Equation \eqref{defSDEY} and that this equation is equivalent to that given in Theorem \ref{ConvergenceTheorem}. The proof of this result can be found in \cite{eleuch2016micro}. Theorem \ref{ConvergenceTheorem} is readily obtained from this proposition together with Proposition \ref{prop1} and \ref{ConvergenceX}.

\begin{prop}
\label{uniqueness}
Let $\lambda$, $\nu$, $\theta$ and $V_0$ be positive constants, $\alpha \in (1/2,1)$ and $B$ be a Brownian motion.
The process $V$ is solution of the following fractional stochastic differential equation
\begin{equation*}
V_t = V_0\big(1-F^{\alpha,\lambda}(t)\big) + \theta F^{\alpha,\lambda}(t) + \nu \int_0^t f^{\alpha,\lambda}(t-s) \sqrt{V_s} dB_s
\end{equation*}
if and only if it is solution of
\begin{equation*}
V_t = V_0 +\frac{1}{\Gamma(\alpha)} \int_0^t (t-s)^{\alpha-1} \lambda (\theta - V_s) ds +  \frac{\lambda \nu}{\Gamma(\alpha)}  \int_0^t (t-s)^{\alpha-1} \sqrt{V_s} dB_s.
\end{equation*}
Furthermore, both equations admit a unique strong solution.
\end{prop}

\subsubsection{Proof of Corollary \ref{ConvergenceHeston}}
From Theorem \ref{ConvergenceTheorem}, we know that $P^T$ converges in law for the Skorokhod topology to the process $P$ given by 
$$ P_t = \sqrt{\frac{\theta}{2}}\int_0^t \sqrt{Y_s} (dB_s^1 - dB_s^2) - \frac{\theta}{2} \int_0^t Y_s ds.$$
Let $V_t= \theta Y_t$ and $W_t = \frac{1}{\sqrt{2}} (B_t^1-B_t^2).$
Then
$$ P_t =  \int_0^t \sqrt{V_s} dW_s - \frac{1}{2} \int_0^t V_s ds, $$
where 
$$ V_t =  \xi \theta + \frac{1}{\Gamma(\alpha)} \int_0^t (t-s)^{\alpha-1} \lambda (\theta-V_s) ds +  \lambda\sqrt{\frac{\theta (1+\beta^2)}{\lambda \mu (1+\beta)^2}} \frac{1}{\Gamma(\alpha)}  \int_0^t (t-s)^{\alpha-1} \sqrt{V_s} dW'_s$$
and $(W,B)$ is a correlated bi-dimensional Brownian motion with
$$ d\langle W,B\rangle_t = \frac{1-\beta}{\sqrt{2(1+\beta^2)}} dt.$$

\subsection{Proof of Theorem \ref{CharacRough}}
\label{proof2}
We now give the proof of Theorem \ref{CharacRough}. We do it for $t\in[0,1]$ but the proof can obviously be extended for any $t\geq 0$. We start by controlling the process $Y^T(a,t)-(1,1)$. In the sequel, $c(a)$ denotes a positive constant independent of $t$ and $T$ that may vary from line to line. 
\subsubsection{Control of $Y^T(a,t)-(1,1)$}
We have the following proposition.
\begin{prop}\label{contrY}
For any $t\in[0,1]$,
$$ T^{\alpha} \|Y^T(a,t) - (1,1)\| \leq c(a).$$
\end{prop}
\noindent\textsc{Proof}:\\
 
\noindent Let us show that
$$T^{\alpha} |Y^{T,+}(a,t) - 1| \leq c(a).$$
Recall that $Y^T(a,t)$ is defined in Section \ref{idee} for $a\in \mathbb R$ by 
$$ Y^T(a,t)= \big(Y^{T,+}(a,t),Y^{T,-}(a,t)\big)= \big{(} C^{T,+}((a_T^+,a_T^-),tT),C^{T,-}((a_T^+,a_T^-),tT) \big{)}, $$
with
$$ a_T^+ =  a \sqrt{\frac{\lambda \theta}{2 \mu}} T^{-\alpha} -  a {\frac{\lambda \theta}{2 \mu}} T^{-2 \alpha},~~a_T^- = - a \sqrt{\frac{\lambda \theta}{2 \mu}} T^{-\alpha}.$$
Using the elements in the proof of Theorem \ref{thHawkes} in Section \ref{demo}, we get that 
$$C^{T,+}\big((a,b),t\big) = \mathbb{E}\big[\exp(ia+ia\tilde{N}_t^{T,+}+ib\tilde{N}_t^{T,-})\big],$$
where $\tilde{N}^{T,} = (\tilde{N}^{T,+},\tilde{N}^{T,-})$ is a bi-dimensional Hawkes process with intensity $(\tilde{\lambda}^{T},\tilde{\lambda}^{T})$ given by
$$\tilde{\lambda}_t^{T} = \frac{1}{\beta+1} \varphi^T(t) + \frac{1}{\beta+1} \int_0^t \varphi^T(t-s) (d\tilde{N}_s^{T,+}+\beta d\tilde{N}_s^{T,-}).$$
As already seen, using Lemma \ref{hopf}, we can rewrite the intensity under the following form:   
$$ \tilde{\lambda}_t^{T} = \frac{1}{\beta+1} \psi^T(t) + \frac{1}{\beta+1} \int_0^t \psi^T(t-s) (d\tilde{M}_s^{T,+}+\beta d\tilde{M}_s^{T,-}),$$
where $\displaystyle\tilde{M}^{T} = (\tilde{M}^{T,+},\tilde{M}^{T,-}) = \tilde{N}^{T} - \int_0^. \tilde{\lambda}^{T}(s)ds(1,1)$ is a martingale. Using Fubini theorem, we get  
$$ \int_0^{tT} \tilde{\lambda}_{s}^{T} ds = \frac{1}{\beta+1} T \int_0^t \psi^T(Ts) ds + \frac{1}{\beta+1} \int_0^t  T \psi^T\big(T(t-s)\big) (\tilde{M}_{sT}^{T,+}+\beta \tilde{M}_{sT}^{T,-}) ds.$$
Then, from \eqref{rho}, we derive
\begin{equation}
 \label{aux1}
 \int_0^{tT} \tilde{\lambda}_{s}^{T} ds = \frac{1}{\lambda (\beta+1)} a_T T^\alpha F^{\alpha,\lambda}(t)+ \frac{1}{\lambda(\beta+1)} a_T T^\alpha \int_0^t  f^{\alpha,\lambda}(t-s) (\tilde{M}_{sT}^{T,+}+\beta \tilde{M}_{sT}^{T,-}) ds. 
\end{equation}
Consequently,
$$ \mathbb{E}\big[\int_0^{tT} \tilde{\lambda}_{s}^{T} ds\big] \leq \frac{1}{\lambda (\beta+1)} F^{\alpha,\lambda}(1)  T^\alpha.$$
Let us now set $\tilde{X}_t^T=a_T^+ \tilde{N}^{T,+}_{tT}+a_T^- \tilde{N}^{T,-}_{tT}$. Using the last inequality, we deduce
$$|\mathbb{E} \tilde{X}_t^T|\leq c|a|T^{-\alpha} F^{\alpha,\lambda}(1).$$
Now recall that
$$T^{\alpha} (Y^{T,+}(a,t) - 1) = T^\alpha\big(\mathbb{E}\big[\exp(ia_T^+ +ia_T^+ \tilde{N}^{T,+}_{tT}+ia_T^- \tilde{N}^{T,-}_{tT})\big] -1\big).$$
Using the fact that there exists $c>0$ such that for any $x \in \mathbb{R}$,
$$|\exp(ix) - 1 - ix | \leq c |x|^2,$$ we obtain 
\begin{align*}
T^{\alpha} |Y^{T,+}(a,t) - 1| &= T^{\alpha} \big|\mathbb{E}\big[\exp(ia_T^+ +i\tilde{X}_t^T) - 1 - i\tilde{X}_t^T- ia_T^++ i\tilde{X}_t^T + i a_T^+\big]\big|\\
&\leq T^{\alpha}|\mathbb{E}[\tilde{X}_t^T]| +T^{\alpha} |a_T^+| + T^{\alpha} \mathbb{E}\big[|\exp(ia_T^+ +i\tilde{X}_t^T) - 1 - i\tilde{X}_t^T- ia_T^+|\big] \\
&\leq c(a) \big(1+ T^{\alpha} (a_T^+)^2 + T^\alpha \mathbb{E}[(\tilde{X}_t^T)^2]\big)\\
&\leq c(a) \big(1+ T^\alpha \mathbb{E}[(\tilde{X}_t^T)^2]\big).
\end{align*}
Then, using that
$$ \tilde{X}_t^T =  a \sqrt{\frac{\lambda \theta}{2 \mu}} T^{-\alpha} (\tilde{N}^{T,+}_{tT}-\tilde{N}^{T,-}_{tT}) - a {\frac{\lambda \theta}{2 \mu}} T^{-2 \alpha} \tilde{N}^{T,+}_{tT}$$
together with the fact that $\tilde{N}^{T,+}-\tilde{N}^{T,-} = \tilde{M}^{T,+}-\tilde{M}^{T,-},$ we deduce
$$
T^{\alpha} \mathbb{E}[(\tilde{X}_t^T)^2]\leq c a^2T^{-\alpha} \mathbb{E}[(\tilde{M}^{T,+}_{tT}-\tilde{M}^{T,-}_{tT})^2] + c a^2 T^{-3\alpha} \mathbb{E}[(\tilde{N}^{T,+}_{tT})^2].
$$
Since $[\tilde{M}^{T,+}-\tilde{M}^{T,-},\tilde{M}^{T,+}-\tilde{M}^{T,-}] = \tilde{N}^{T,+}+\tilde{N}^{T,-},$ we get
\begin{align*}
T^{\alpha} \mathbb{E}[(\tilde{X}_t^T)^2] &\leq c a^2 T^{-\alpha} \mathbb{E}[\tilde{N}^{T,+}_{tT}+\tilde{N}^{T,-}_{tT}] + c a^2 T^{-3\alpha} \mathbb{E}[(\tilde{N}^{T,+}_{tT})^2]\\
&\leq c a^2 \big( T^{-\alpha} \mathbb{E}[\int_0^{tT} \tilde{\lambda}_s^{T}ds] +  T^{-3\alpha} \mathbb{E}[(\tilde{N}^{T,+}_{tT})^2]\big).\\
&\leq c a^2\big(1+ T^{-3\alpha} \mathbb{E}[(\tilde{N}^{T,+}_{tT})^2]\big).
\end{align*}
In order to control the term $\mathbb{E}[(\tilde{N}^{T,+}_{tT})^2]$, we now compute a bound for $\displaystyle\mathbb{E}\big[(\int_0^{tT}\tilde{\lambda}_s^T ds)^2\big]$. Using \eqref{aux1}, this last quantity is equal to
$$\frac{1}{\lambda^2 (\beta+1)^2} a_T^2 T^{2\alpha} \big(F^{\alpha,\lambda}(t)\big)^2+ \frac{1}{\lambda^2(\beta+1)^2} a_T^2 T^{2\alpha} \mathbb{E} \Big[\Big(\int_0^t  f^{\alpha,\lambda}(t-s) (\tilde{M}_{sT}^{T,+}+\beta \tilde{M}_{sT}^{T,-}) ds \Big)^2\Big],$$
which is smaller than
$$c(a) T^{2 \alpha}\Big( 1 + \mathbb{E} \big[\int_0^t \big(f^{\alpha,\lambda}(t-s)\big)^2 (\tilde{M}_{sT}^{T,+}+\beta \tilde{M}_{sT}^{T,-})^2 ds \big]\Big).$$
Since $[\tilde{M}^{T,+}+\beta \tilde{M}^{T,-},\tilde{M}^{T,+} + \beta \tilde{M}^{T,-}] = \tilde{N}^{T,+}+ \beta^2 \tilde{N}^{T,-}$, we obtain 
\begin{align*}
\mathbb{E}\big[(\int_0^{tT}\tilde{\lambda}_s^T ds)^2\big] &\leq c(a) T^{2 \alpha}\Big( 1 + \int_0^t  \big(f^{\alpha,\lambda}(t-s)\big)^2 \mathbb{E}[ \tilde{N}_{sT}^{T,+}+\beta^2 \tilde{N}_{sT}^{T,-}] ds \Big)\\
&\leq c(a) T^{2 \alpha}\Big(1 + \int_0^t  \big(f^{\alpha,\lambda}(t-s)\big)^2 \mathbb{E}\big[\int_0^{sT} \tilde{\lambda}_{u}^{T} du\big] ds \Big)\\
&\leq c(a) T^{2 \alpha}\Big(1 + T^{\alpha} \int_0^1  \big(f^{\alpha,\lambda}(s)\big)^2 ds  \Big)\\
&\leq c(a) T^{3 \alpha}.
\end{align*}
Thus 
$$
\mathbb{E}\big[(\tilde{N}^{T,+}_{tT})^2\big] \leq 2 \mathbb{E}\big[(\tilde{M}^{T,+}_{tT})^2\big] + 2 \mathbb{E}\big[\big(\int_0^{tT} \tilde{\lambda}_{s}^{T} ds \big)^2\big]\leq c(a) T^{3 \alpha}.
$$
Finally, $ T^{\alpha} \mathbb{E}[(\tilde{X}_t^T)^2] \leq c(a)$ and therefore
$$T^{\alpha} |Y^{T,+}(a,t) - 1| \leq c(a).$$ The fact that
$$T^{\alpha} |Y^{T,-}(a,t) - 1| \leq c(a) $$ is proved similarly.
\qed
\subsubsection{Convergence of $T^{\alpha}(Y^{T}-(1,1))$}
Let $\kappa=\lambda \theta/(2 \mu)$. We have the following proposition.
\begin{prop}\label{propconvY}
The sequence $T^\alpha(Y^T(a,t)-(1,1))$ converges uniformly in $t \in [0,1]$ to $\big(c(a,t),d(a,t)\big)$, where $(c,d)$ are solutions of
$$ c(a,t) =  ia\sqrt{\kappa} - ia \frac{\kappa}{\lambda(\beta+1)}  F^{\alpha,\lambda}(t)  + \frac{1}{2\lambda(\beta+1)}  \int_0^t \big(c^2(a,t-s) +d^2(a,t-s)\big) f^{\alpha,\lambda}(s)ds  $$
$$ d(a,t) =  - ia\sqrt{\kappa} - ia \frac{\beta \kappa}{\lambda(\beta+1)}  F^{\alpha,\lambda}(t)  + \frac{\beta}{2\lambda(\beta+1)}  \int_0^t \big(c^2(a,t-s) +d^2(a,t-s)\big) f^{\alpha,\lambda}(s)ds.$$
\end{prop}
\noindent\textsc{Proof}: 
\paragraph{Convenient rewriting of $T^{\alpha}(Y^{T}-(1,1))$}
Using the fact that the complex logarithm is analytic on the set $\mathbb C/\mathbb R^-$, we can show that there exists $c > 0$ such that for any $x \in \mathbb{C}$ with $|x| <1/2$, 
$$|\log(1+x) - x + \frac{1}{2} x^2| \leq c |x|^3.$$ 
Thus we can write
$$ \log\big(Y^T(a,t)\big) = Y^T(a,t) - (1,1) - \frac{1}{2} \big(Y^T(a,t) - (1,1)\big)^2 - \varepsilon^T(a,t),$$
with $|\varepsilon^T(a,t)| \leq c(a)T^{-3\alpha}$. Indeed, for large enough $T$, we have from Proposition \ref{contrY} that $|Y^{T,+}(a,t)-1| \leq 1/2$ and $|Y^{T,-}(a,t)-1| \leq 1/2$, uniformly in $t$. Now, again from Proposition \ref{contrY}, it is easy to see that
$$ \big\|i(a_T^+ , a_T^- ) + \int_0^tT\big(Y^T(a, t - s) - (1, 1)\big).\phi^T (Ts)ds\big\| \leq c(a) T^{-\alpha} \underset{T \rightarrow \infty}{\longrightarrow} 0.$$ 
Hence, for large enough $T$, the imaginary part of $$i(a_T^+ , a_T^- ) + \int_0^t T(Y^T(a, t - s) - (1, 1)).\phi^T (Ts)ds$$
has a norm which is smaller than $\pi$. Therefore 
$$ \log\Big(\exp\big(i(a_T^+ , a_T^- ) + \int_0^tT\big(Y^T(a, t - s) - (1, 1)\big).\phi^T (Ts)ds\big)\Big)$$
is equal to  $$i(a_T^+ , a_T^- ) + \int_0^tT\big(Y^T(a, t - s) - (1, 1)\big).\phi^T (Ts)ds.$$
Then, using Equation \eqref{decY}, we get
\begin{align*}
Y^T(a,t) - (1,1) &= \frac{1}{2} \big(Y^T(a,t) - (1,1)\big)^2 + \varepsilon^T(a,t) + ia\sqrt{\kappa}T^{-\alpha}(1,-1) \\
&-i a \kappa T^{-2\alpha}(1,0) +T \int_0^t\big(Y^T(a,t-s) - (1,1)\big).\phi^T(Ts)ds.
\end{align*}
Using again the fact that
$$
\sum_{k \geq 1} \big(T \phi^T(T.) \big)^{*k}= a_T \frac{T^{\alpha}}{\lambda} f^{\alpha,\lambda} \chi,$$
together with Lemma \ref{hopf}, we derive 
\begin{align*}
Y^T(a,t) - (1,1) &= \frac{1}{2} \big(Y^T(a,t) - (1,1)\big)^2 + \varepsilon^T(a,t) + ia\sqrt{\kappa}T^{-\alpha}(1,-1) - ia \kappa T^{-2\alpha}(1,0)  \\
&+\frac{a_T}{2} \frac{T^{\alpha}}{\lambda} \int_0^t \big(Y^T(a,t-s) - (1,1)\big)^2.\chi f^{\alpha,\lambda}(s)ds +   \frac{a_T}{\lambda} T^{\alpha}\int_0^t \varepsilon^T(a,t-s).\chi f^{\alpha,\lambda}(s)ds \\
&+ ia \sqrt{\kappa}\frac{a_T}{\lambda}  (1,-1).\chi F^{\alpha,\lambda}(t) - ia \kappa T^{-\alpha} \frac{a_T}{\lambda} (1,0).\chi F^{\alpha,\lambda}(t).
\end{align*}
Let $$ \varepsilon_1^T(a,t)  = \frac{1}{2} \big(Y^T(a,t) - (1,1)\big)^2 + \varepsilon^T(a,t) - ia \kappa T^{-2\alpha}(1,0)  + \frac{a_T}{\lambda} T^{\alpha} \int_0^t \varepsilon^T(a,t-s).\chi f^{\alpha,\lambda}(s)ds.$$ We have
 \begin{align*}
Y^T(a,t) - (1,1) &= \varepsilon_1^T(a,t) + ia\sqrt{\kappa}T^{-\alpha}(1,-1) + \frac{a_T}{2} \frac{T^{\alpha}}{\lambda}  \int_0^t \big(Y^T(a,t-s) - (1,1)\big)^2.\chi f^{\alpha,\lambda}(s)ds \\
& - ia_T a \frac{\kappa}{\lambda(\beta+1)} T^{-\alpha} F^{\alpha,\lambda}(t) (1,\beta).
\end{align*}
Let now $$ \varepsilon_2^T(a,t)  = - \frac{1}{2}  \int_0^t \big(Y^T(a,t-s) - (1,1)\big)^2.\chi f^{\alpha,\lambda}(s)ds + ia \frac{\kappa}{(\beta+1)} T^{-2\alpha} F^{\alpha,\lambda}(t) (1,\beta).$$ We obtain
\begin{align*}
Y^T(a,t) - (1,1)&= \varepsilon_1^T(a,t) + \varepsilon_2^T(a,t) + ia\sqrt{\kappa}T^{-\alpha}(1,-1) + \frac{1}{2\lambda} T^\alpha \int_0^t \big(Y^T(a,t-s) - (1,1)\big)^2.\chi f^{\alpha,\lambda}(s)ds \\
& - ia \frac{\kappa}{\lambda(\beta+1)} T^{-\alpha} F^{\alpha,\lambda}(t) (1,\beta).
\end{align*}
Using Proposition \ref{contrY}, we easily see that $T^{2\alpha} \varepsilon_1^T$ and $T^{2\alpha} \varepsilon_2^T$ are uniformly bounded in $t$ and $T$. We now set
$$ \theta^T(a,t)= \big(\theta^{T,+}(a,t),\theta^{T,-}(a,t)\big)=  T^\alpha\big(Y^T(a,t) - (1,1)\big) $$ and
$$ r^T(a,t)= T^{\alpha} \big(\varepsilon_1^T(a,t) + \varepsilon_2^T(a,t)\big).$$
We have that $T^{\alpha}r^T$ is uniformly bounded in $t$ and $T$ and
$$ \theta^T(a,t) = r^T(a,t) + i a\sqrt{\kappa}(1,-1) - ia \frac{\kappa}{\lambda(\beta+1)}  F^{\alpha,\lambda}(t) (1,\beta) + \frac{1}{2\lambda}  \int_0^t \big(\theta^T(a,t-s)\big)^2.\chi f^{\alpha,\lambda}(s)ds.$$ 
\paragraph{Convergence of $\theta^T$}
For fixed $a$, we now show that $t \rightarrow \theta^T(a,t) $ is a Cauchy sequence in the space of continuous functions $C([0,1],\mathbb{R}^2)$ equipped with the sup-norm. Let $\delta>0$ and $T_0>1$ such that for $ T>T_0 $ , $\|r^T(a,t)\|_{\infty}\leq \frac{\delta}{2}$ for any $t\in [0,1]$. Then for $T> T_0$, $T' > T_0$ and $t\in [0,1]$,
$$ \|\theta^T(a,t) - \theta^{T'}(a,t)\|\leq \delta + \frac{1}{2\lambda}  \int_0^t \big\|\big(\theta^T(a,t-s)\big)^2.\chi  - \big(\theta^{T'}(a,t-s)\big)^2.\chi \big\| f^{\alpha,\lambda}(s)ds.$$
Since $\theta^T$ is uniformly bounded in $t$ and $T$, we get
$$\|\theta^T(a,t) - \theta^{T'}(a,t)\| \leq\delta + C(a)  \int_0^t \|\theta^T(a,t-s)  - \theta^{T'}(a,t-s) \| f^{\alpha,\lambda}(s)ds.$$
Using Lemma \ref{UsefulIneq} in Appendix, this enables us to show that $\theta^T$ is a Cauchy sequence. Consequently, $\theta^T(a,t)$ converges uniformly in $t$ to $\big(c(a,t),d(a,t)\big)$, where $(c,d)$ is solution to the following equation:
$$ c(a,t) =  ia\sqrt{\kappa} - ia \frac{\kappa}{\lambda(\beta+1)}  F^{\alpha,\lambda}(t)  + \frac{1}{2\lambda(\beta+1)}  \int_0^t \big(c^2(a,t-s) +d^2(a,t-s)\big) f^{\alpha,\lambda}(s)ds  $$
$$ d(a,t) =  - ia\sqrt{\kappa} - ia \frac{\beta \kappa}{\lambda(\beta+1)}  F^{\alpha,\lambda}(t)  + \frac{\beta}{2\lambda(\beta+1)}  \int_0^t \big(c^2(a,t-s) +d^2(a,t-s)\big) f^{\alpha,\lambda}(s)ds.$$\qed
\subsubsection{End of the proof of Theorem \ref{CharacRough}}
\paragraph{Deriving the characteristic function}
Let $a\in\mathbb{R}$. Recall that from Section \ref{idee}, we have
$$ L^T(a_T^+,a_T^-,tT) = \exp \Big(\int_0^t \big(T^\alpha (Y^{T,+}(a,t-s) -1 ) + T^\alpha (Y^{T,-}(a,t-s) -1 )\big) \big(T^{1-\alpha} \hat{\mu}(sT)\big) ds \Big) $$
and furthermore, from Proposition \ref{propconvY},
$$T^\alpha (Y^{T,+}(a,t) -1 ) + T^\alpha (Y^{T,-}(a,t) -1 ) $$
converges uniformly in $t$ to $c(a,t)+d(a,t)$.  
Also, using Remark \ref{mu}, we have 
$$ T^{1-\alpha} \hat{\mu}(tT) = \mu + \mu \xi \big(\frac{t^{-\alpha}}{\lambda} (Tt)^\alpha \int_{tT}^\infty \varphi(s) ds + \lambda T^{-\alpha} \int_0^{tT} \varphi(s) ds \big)$$
and therefore $T^{1-\alpha} \hat{\mu}(tT)$ converges towards 
$$ \mu\big(1+  \xi \frac{t^{-\alpha}}{\lambda \Gamma(1-\alpha)}\big).$$
In addition, using Proposition \ref{contrY}, we get that for given $t \in [0,1]$ and for any $s \in [0,t]$ 
$$ \big| T^\alpha (Y^{T,+}(a,t-s) -1 ) + T^\alpha (Y^{T,-}(a,t-s) -1 )  \big|\big(T^{1-\alpha} \hat{\mu}(sT)\big)\leq c(a)(1+s^{-\alpha}).$$ The right hand side of the last inequality is integrable over $[0,t]$.
Therefore, using the convergence of $L^T(a_T^+,a_T^-,tT)$ towards $L_p(a,t)$ and applying the dominated convergence theorem, we obtain
$$ L_p(a,t) =   \exp \big( \int_0^t g(a,s) (1+ \xi \frac{(t-s)^{-\alpha}}{\lambda \Gamma(1-\alpha)})ds \big),$$
where $g(a,t) = \mu \big(c(a,t) +d(a,t)\big) $. Thus, we have shown that
$$ L_p(a,t) = \exp \big{(} \int_0^t g(a,s) ds + \frac{V_0}{\theta \lambda} I^{1-\alpha}g(a,t)  \big{)}.$$
\paragraph{Integral equation for $g$}
We now prove that $g$ is solution of an integral equation. First remark that
$$ d(a,t) = \beta c(a,t) - ia (1+\beta) \sqrt{\kappa}.$$
Hence $g(a,t)=\mu (\beta +1) (c(a,t)- ia \sqrt{\kappa})$, which can be written
$$- ia \frac{\mu \kappa}{\lambda} F^{\alpha,\lambda}(t) + \frac{\mu}{2\lambda}  \int_0^t \Big((c(a,s)-ia\sqrt{\kappa} + ia\sqrt{\kappa})^2 + \big(\beta(c(a,s)-ia\sqrt{\kappa}) - ia\sqrt{\kappa}\big)^2\Big) f^{\alpha,\lambda}(t-s) ds.$$
Thus,
\begin{align*}
g(a,t)&= - ia \frac{\mu \kappa}{\lambda} F^{\alpha,\lambda}(t) + \frac{1+\beta^2}{2 \mu \lambda(1+\beta)^2}  \int_0^t \big(g(a,s)\big)^2 f^{\alpha,\lambda}(t-s) ds - a^2  \frac{\mu \kappa}{\lambda}F^{\alpha,\lambda}(t) \\
&+  ia \frac{\sqrt{\kappa}(1-\beta)}{\lambda(\beta+1)}  \int_0^t g(a,s) f^{\alpha,\lambda}(t-s) ds.
\end{align*}
Using the definition of $\kappa$ in Section \ref{proof2}, we deduce
\begin{align*}
g(a,t)&=  \frac{\theta}{2}(-a^2 - ia) F^{\alpha,\lambda}(t) +  ia \frac{\sqrt{\theta}(1-\beta)}{\sqrt{2\lambda\mu}(\beta+1)}  \int_0^t g(a,s) f^{\alpha,\lambda}(t-s) ds\\ 
&+\frac{1+\beta^2}{2\mu \lambda(1+\beta)^2}  \int_0^t g^2(a,s) f^{\alpha,\lambda}(t-s) ds
\end{align*}
and from those of $\rho$ and $\nu$ in Section \ref{idee}, we finally obtain that $g(a,t)$ is equal to
$$\frac{\theta}{2}(-a^2 - ia) F^{\alpha,\lambda}(t) +  ia \rho \nu \int_0^t g(a,s) f^{\alpha,\lambda}(t-s) ds +\frac{\nu^2}{2 \theta}  \int_0^t \big(g(a,s)\big)^2 f^{\alpha,\lambda}(t-s) ds.$$ Thus,
$$ L_p(a,t) =\exp \Big{(} \int_0^t g(a,s) \big(1+ \xi \frac{(t-s)^{-\alpha}}{\lambda \Gamma(1-\alpha)}\big)ds \Big{)} $$
with
$$g(a,t) = \int_0^t \Big(\frac{\theta}{2}(-a^2 - ia)+ ia \rho \nu g(a,s) + \frac{\nu^2}{2\theta}\big(g(a,s)\big)^2\Big) f^{\alpha,\lambda}(t-s) ds.$$
Let us now set $h =g/(\theta \lambda)$. Then
$$ L_p(a,t) =\exp \Big{(} \int_0^t h(a,s) \big(\theta \lambda+ V_0 \frac{(t-s)^{-\alpha}}{ \Gamma(1-\alpha)}\big)ds \Big{)},$$
with 
\begin{equation}
\label{FracEDP2}
h(a,t) = \int_0^t \Big(\frac{1}{2}(-a^2 - ia)+ ia \lambda \rho \nu h(a,s) + \frac{(\lambda \nu)^2}{2}\big(h(a,s)\big)^2\Big) \frac{1}{\lambda}f^{\alpha,\lambda}(t-s) ds.
\end{equation}
Using Lemma \ref{EDPFracLin}, we have that Equation \eqref{FracEDP2} can also be written under the following form:
$$ D^\alpha h(a,t) = \frac{1}{2}(-a^2-ia)+ \lambda (i a \rho \nu-1)  h(a,s) + \frac{(\lambda \nu)^2}{2 } \big(h(a,s)\big)^2,~~I^{1-\alpha}h(a,0)=0.$$
\subsubsection{Uniqueness of the solution of \eqref{FracEDP}}
For a given $a \in \mathbb{R}$, consider two continuous solutions $h_1(a,.)$ and $h_2(a,.)$ of \eqref{FracEDP} or equivalently of \eqref{FracEDP2}. We have that
$|h_1(a,t) - h_2(a,t)|$ is smaller than
$$\int_0^t \big( |a \rho \nu|  |h_1(a,s) - h_2(a,s)| + \frac{\lambda \nu^2}{2}|\big(h_1(a,s)\big)^2 - \big(h_2(a,s)\big)^2| \big) f^{\alpha,\lambda}(t-s) ds.$$
Using the continuity of $h_1(a,.)$ and $h_2(a,.)$, this is also smaller than
$$c(a) \int_0^t |h_1(a,s) - h_2(a,s)| f^{\alpha,\lambda}(t-s) ds.$$
Thanks to Lemma \ref{UsefulIneq}, this gives $h_1(a,.) = h_2(a,.)$.

\section*{Acknowledgments}

We thank Masaaki Fukasawa, Jim Gatheral and Antoine Jacquier for many interesting discussions and Christa Cuchiero and Josef Teichmann for very relevant comments about the affine nature of the processes considered in this work.

\appendix

\section{Appendix}
We gather in this section some useful technical results.
\subsection{Mittag-Leffler functions}
\label{mittag}
Let $(\alpha,\beta) \in (\mathbb{R}_+^*)^2$. The Mittag-Leffler function $E_{\alpha,\beta}$ is defined for $z \in \mathbb{C}$ by
$$ E_{\alpha,\beta}(z) = \sum_{n \geq 0} \frac{z^n}{\Gamma(\alpha n + \beta)}.$$ 
For $(\alpha,\lambda) \in  (0,1)\times \mathbb{R}_+$, we also define 
$$ f^{\alpha,\lambda}(t) = \lambda t^{\alpha - 1} E_{\alpha,\alpha}(-\lambda t^\alpha),~~t>0,$$
$$ F^{\alpha,\lambda} = \int_0^t f^{\alpha,\lambda}(s) ds,~~t \geq 0.$$
The function $f^{\alpha,\lambda}$ is a density function on $\mathbb{R}_+$ called Mittag-Leffler density function.
The following properties of $f^{\alpha,\lambda}$ and $F^{\alpha,\lambda}$ can be found in \cite{haubold2011mittag,mainardisome,mathai2008special}.
We have
$$ f^{\alpha,\lambda}(t) \underset{t \rightarrow 0^+}{\sim} \frac{\lambda}{\Gamma(\alpha)} t^{\alpha-1},~~f^{\alpha,\lambda}(t) \underset{t \rightarrow \infty}{\sim} \frac{\alpha}{\lambda \Gamma(1-\alpha)} t^{-(\alpha+1)}$$
and
$$ F^{\alpha,\lambda}(t) = 1 - E_{\alpha,1}(-\lambda t^\alpha),~~F^{\alpha,\lambda}(t) \underset{t \rightarrow 0^+}{\sim} \frac{\lambda}{\Gamma(\alpha+1)} t^{\alpha},~~ 1-F^{\alpha,\lambda}(t) \underset{t \rightarrow \infty}{\sim} \frac{1}{\lambda \Gamma(1-\alpha)} t^{-\alpha}.$$
Finally, for $\alpha \in (1/2,1)$, $f^{\alpha,\lambda}$ is square-integrable and its Laplace transform is given for $z \geq 0$ by 
$$\hat{f}^{\alpha,\lambda}(z)= \int_0^\infty f_{\alpha,\lambda}(s) e^{-zs} ds = \frac{\lambda}{\lambda + z^\alpha}.$$

\subsection{Wiener-Hopf equations}
The following result is used extensively in this work to solve Wiener-Hopf type equations, see for example \cite{bacry2013some}.
\begin{lemma}
\label{hopf}
Let $g$ be a measurable locally bounded function from $\mathbb{R}$ to $\mathbb{R}^d$ and $\phi : \mathbb{R}_+ \rightarrow \cal{M}^{\textbf{d}}(\mathbb{R}) $ be a matrix-valued function with integrable components such that $\mathcal{S}(\int_0^\infty \phi(s) ds) < 1$. Then there exists a unique
locally bounded function $f$ from $\mathbb{R}$ to $\mathbb{R}^d$ solution of
$$f(t) = g(t) + \int_0^t \phi(t-s). f(s) ds,~~t \geq 0$$
given by
$$ f(t) = g(t) + \int_0^t \psi(t-s). g(s) ds,~~t \geq 0, $$
where $\displaystyle\psi= \sum_{k \geq 1} \phi^{*k}$.
\end{lemma}

\subsection{\textsc{Fractional differential equations}}
We end this appendix with some useful results about fractional differential equations. The next lemma can be found in \cite{samko1993fractional}.
\begin{lemma}
\label{EDPFracLin} 
Let $h$ be a continuous function from $[0,1]$ to $\mathbb{R}$, $\alpha \in (0,1]$ and $\lambda\in\mathbb{R}$. There is a unique continuous solution to the equation
\begin{equation*}
D^\alpha y(t) = \lambda y(t) + h(t),~~I^{1-\alpha}y(0) = 0
\end{equation*}
given by
$$ y(t) = \int_0^t (t-s)^{\alpha-1} E_{\alpha,\alpha}\big(\lambda(t-s)^\alpha\big) h(s) ds.$$
\end{lemma}
\noindent We also have the following useful result.
\begin{lemma}
\label{UsefulIneq}
Let $h$ be a non-negative continuous function from $[0,1]$ to $\mathbb{R}$ such that for any $t \in[0,1]$,
$$ h(t) \leq \varepsilon + C \int_0^t f^{\alpha,\lambda}(t-s) h(s) ds,$$
for some $\varepsilon\geq 0$ and $C \geq 0$. Then for any $t \in[0,1]$, 
$$ h(t) \leq C' \varepsilon,$$
with $$C' = 1+ C\lambda \int_0^{1} s^{\alpha-1} E_{\alpha,\alpha}\big(\lambda(C-1)s^\alpha\big) ds>0.$$
In particular, if $\varepsilon =0$ then $h = 0$.
\end{lemma}
\noindent\textsc{Proof}:\\

\noindent Let $$f(t)=h(t) -  C \int_0^t f^{\alpha,\lambda}(t-s) h(s) ds.$$ and $g=h-f$. The function $g$ is solution of 
$$g(t) = C \int_0^t f^{\alpha,\lambda}(t-s) \big(g(s)+f(s)\big) ds.$$
Thus, from Lemma \ref{EDPFracLin}, $g$ is the unique solution of
$$ D^\alpha g(t) = \lambda(C-1) g(t) + C\lambda f(t),~~I^{1-\alpha}g(0)=0.$$
Hence using again Lemma \ref{EDPFracLin}, we deduce that 
$$g(t) = C\lambda \int_0^t (t-s)^{\alpha-1} E_{\alpha,\alpha}\big(\lambda(C-1)(t-s)^\alpha\big) f(s) ds.$$
Therefore,
$$ g(t) \leq C\lambda \varepsilon \int_0^t s^{\alpha-1} E_{\alpha,\alpha}\big(\lambda(C-1)s^\alpha\big)ds.$$
Using that  $h=f+g$ together with the fact that $E_{\alpha,\alpha}$ is non-negative, we get the result.
\qed

\bibliographystyle{abbrv}
\bibliography{BibEER2_final}

\begin{thebibliography}{10}

\bibitem{albrecher2006little}
H.~Albrecher, P.~Mayer, W.~Schoutens, and J.~Tistaert.
\newblock The little {H}eston trap.
\newblock {\em Wilmott Magazine}, pages 83--92, January 2007.

\bibitem{bacry2013modelling}
E.~Bacry, S.~Delattre, M.~Hoffmann, and J.-F. Muzy.
\newblock Modelling microstructure noise with mutually exciting point
  processes.
\newblock {\em Quantitative Finance}, 13(1):65--77, 2013.

\bibitem{bacry2013some}
E.~Bacry, S.~Delattre, M.~Hoffmann, and J.-F. Muzy.
\newblock Some limit theorems for {H}awkes processes and application to
  financial statistics.
\newblock {\em Stochastic Processes and their Applications}, 123(7):2475--2499,
  2013.

\bibitem{bacry2014estimation}
E.~Bacry, T.~Jaisson, and J.-F. Muzy.
\newblock Estimation of slowly decreasing {H}awkes kernels: Application to high
  frequency order book modelling.
\newblock {\em Quantitative Finance}, 16(8):1179--1201, 2016.

\bibitem{bayer2016pricing}
C.~Bayer, P.~Friz, and J.~Gatheral.
\newblock Pricing under rough volatility.
\newblock {\em Quantitative Finance}, 16(6):887--904, 2016.

\bibitem{bennedsen2015hybrid}
M.~Bennedsen, A.~Lunde, and M.~S. Pakkanen.
\newblock Hybrid scheme for {B}rownian semistationary processes.
\newblock {\em arXiv preprint arXiv:1507.03004}, 2015.

\bibitem{bouchaud2003theory}
J.-P. Bouchaud and M.~Potters.
\newblock {\em Theory of financial risk and derivative pricing: from
  statistical physics to risk management}.
\newblock Cambridge university press, 2003.

\bibitem{carr1999option}
P.~Carr and D.~Madan.
\newblock Option valuation using the fast {F}ourier transform.
\newblock {\em Journal of Computational Finance}, 2(4):61--73, 1999.

\bibitem{christie1982stochastic}
A.~A. Christie.
\newblock The stochastic behavior of common stock variances: Value, leverage
  and interest rate effects.
\newblock {\em Journal of Financial Economics}, 10(4):407--432, 1982.

\bibitem{diethelm2002predictor}
K.~Diethelm, N.~J. Ford, and A.~D. Freed.
\newblock A predictor-corrector approach for the numerical solution of
  fractional differential equations.
\newblock {\em Nonlinear Dynamics}, 29(1-4):3--22, 2002.

\bibitem{diethelm2004detailed}
K.~Diethelm, N.~J. Ford, and A.~D. Freed.
\newblock Detailed error analysis for a fractional {A}dams method.
\newblock {\em Numerical algorithms}, 36(1):31--52, 2004.

\bibitem{diethelm1998fracpece}
K.~Diethelm and A.~D. Freed.
\newblock The fracpece subroutine for the numerical solution of differential
  equations of fractional order.
\newblock In {\em Forschung und Wissenschaftliches Rechnen 1998}, pages 57--71.
  Gesellschaft f{\"u}r Wisseschaftliche Datenverarbeitung Gottingen, Germany,
  1999.

\bibitem{dragulescu2002probability}
A.~A. Dragulescu and V.~M. Yakovenko.
\newblock Probability distribution of returns in the {H}eston model with
  stochastic volatility.
\newblock {\em Quantitative finance}, 2(6):443--453, 2002.

\bibitem{eleuch2016micro}
O.~El~Euch, M.~Fukasawa, and M.~Rosenbaum.
\newblock The microstructural foundations of leverage effect and rough
  volatility.
\newblock {\em Working paper}, 2016.

\bibitem{forde2012small}
M.~Forde, A.~Jacquier, and R.~Lee.
\newblock The small-time smile and term structure of implied volatility under
  the {H}eston model.
\newblock {\em SIAM Journal on Financial Mathematics}, 3(1):690--708, 2012.

\bibitem{fukasawa2011asymptotic}
M.~Fukasawa.
\newblock Asymptotic analysis for stochastic volatility: {M}artingale
  expansion.
\newblock {\em Finance and Stochastics}, 15(4):635--654, 2011.

\bibitem{gatheral2011volatility}
J.~Gatheral.
\newblock {\em The volatility surface: a practitioner's guide}, volume 357.
\newblock John Wiley \& Sons, 2011.

\bibitem{gatheral2014volatility}
J.~Gatheral, T.~Jaisson, and M.~Rosenbaum.
\newblock Volatility is rough.
\newblock {\em Available at SSRN 2509457}, 2014.

\bibitem{guennoun2014asymptotic}
H.~Guennoun, A.~Jacquier, and P.~Roome.
\newblock Asymptotic behaviour of the fractional {H}eston model.
\newblock {\em Available at SSRN 2531468}, 2014.

\bibitem{hardiman2013critical}
S.~J. Hardiman, N.~Bercot, and J.-P. Bouchaud.
\newblock Critical reflexivity in financial markets: a {H}awkes process
  analysis.
\newblock {\em The European Physical Journal B}, 86(10):1--9, 2013.

\bibitem{haubold2011mittag}
H.~J. Haubold, A.~M. Mathai, and R.~K. Saxena.
\newblock Mittag-leffler functions and their applications.
\newblock {\em Journal of Applied Mathematics}, 2011.

\bibitem{hawkes1974cluster}
A.~G. Hawkes and D.~Oakes.
\newblock A cluster process representation of a self-exciting process.
\newblock {\em Journal of Applied Probability}, pages 493--503, 1974.

\bibitem{heston1993closed}
S.~L. Heston.
\newblock A closed-form solution for options with stochastic volatility with
  applications to bond and currency options.
\newblock {\em Review of Financial Studies}, 6(2):327--343, 1993.

\bibitem{itkin2005pricing}
A.~Itkin.
\newblock Pricing options with {VG} model using {FFT}.
\newblock {\em arXiv preprint physics/0503137}, 2005.

\bibitem{jacod2013limit}
J.~Jacod and A.~Shiryaev.
\newblock {\em Limit theorems for stochastic processes}, volume 288.
\newblock Springer Science \& Business Media, 2013.

\bibitem{jacquier2013small}
A.~Jacquier and P.~Roome.
\newblock The small-maturity {H}eston forward smile.
\newblock {\em SIAM Journal on Financial Mathematics}, 4(1):831--856, 2013.

\bibitem{jacquier2016large}
A.~Jacquier and P.~Roome.
\newblock Large-maturity regimes of the {H}eston forward smile.
\newblock {\em Stochastic Processes and their Applications}, 126(4):1087--1123,
  2016.

\bibitem{jaisson2015rough}
T.~Jaisson and M.~Rosenbaum.
\newblock Rough fractional diffusions as scaling limits of nearly unstable
  heavy tailed {H}awkes processes.
\newblock {\em The Annals of Applied Probability}, to appear, 2016.

\bibitem{jaisson2015limit}
T.~Jaisson, M.~Rosenbaum, et~al.
\newblock Limit theorems for nearly unstable {H}awkes processes.
\newblock {\em The Annals of Applied Probability}, 25(2):600--631, 2015.

\bibitem{janek2011fx}
A.~Janek, T.~Kluge, R.~Weron, and U.~Wystup.
\newblock {FX} smile in the {H}eston model.
\newblock In {\em Statistical Tools for Finance and Insurance}, pages 133--162.
  Springer, 2011.

\bibitem{kahl2005not}
C.~Kahl and P.~J{\"a}ckel.
\newblock Not-so-complex logarithms in the {H}eston model.
\newblock {\em Wilmott magazine}, pages 94--103, September 2005.

\bibitem{lewis2001simple}
A.~L. Lewis.
\newblock A simple option formula for general jump-diffusion and other
  exponential l{\'e}vy processes.
\newblock {\em Available at SSRN 282110}, 2001.

\bibitem{li2009fractional}
C.~Li and C.~Tao.
\newblock On the fractional {A}dams method.
\newblock {\em Computers \& Mathematics with Applications}, 58(8):1573--1588,
  2009.

\bibitem{mainardisome}
F.~Mainardi.
\newblock On some properties of the {M}ittag-{L}effler function.
\newblock {\em arXiv preprint arXiv:1305.0161}.

\bibitem{mandelbrot1997variation}
B.~B. Mandelbrot.
\newblock The variation of certain speculative prices.
\newblock In {\em Fractals and Scaling in Finance}, pages 371--418. Springer,
  1997.

\bibitem{mathai2008special}
A.~M. Mathai and H.~J. Haubold.
\newblock {\em Special functions for applied scientists}.
\newblock Springer, 2008.

\bibitem{mazzon2015forward}
A.~Mazzon and A.~Pascucci.
\newblock The forward smile in local-stochastic volatility models.
\newblock {\em Available at SSRN 2560300}, 2015.

\bibitem{poon2009heston}
S.-H. Poon.
\newblock The {H}eston option pricing model.
\newblock {\em Unpublished Draft}, 2009.

\bibitem{revuz1999continuous}
D.~Revuz and M.~Yor.
\newblock {\em Continuous martingales and Brownian motion}, volume 293.
\newblock Springer Science \& Business Media, 1999.

\bibitem{samko1993fractional}
S.~G. Samko, A.~A. Kilbas, and O.~I. Marichev.
\newblock {\em Fractional integrals and derivatives}, volume 1993.
\newblock Theory and Applications, Gordon and Breach, Yverdon, 1993.

\bibitem{schmelzle2010option}
M.~Schmelzle.
\newblock Option pricing formulae using {F}ourier transform: Theory and
  application.
\newblock {\em Preprint, http://pfadintegral. com}, 2010.

\end{thebibliography}

\end{document}